\documentclass[a4paper,fleqn,usenatbib]{mnras}
\usepackage[T1]{fontenc}
\usepackage{ae,aecompl}
\usepackage{longtable}
\usepackage{makecell}
\usepackage{multirow}
\usepackage{supertabular,booktabs}
\usepackage{upgreek}

\usepackage{graphicx}	
\usepackage{amssymb}	

\usepackage{url}
\usepackage{color}
\usepackage{url}
\usepackage{journals}
\usepackage{array,longtable}

\newcommand {\be}{\begin {equation}}
\newcommand {\ee}{\end {equation}}

\newcommand{\aql}{Aql~X-1}
\newcommand{\lr}{$L_\mathrm{R}$}
\newcommand{\lx}{$L_\mathrm{X}$}
\newcommand{\thh}{$^{\mathrm{th}}$}

\newcommand{\detwo}{$0.34^{+0.29}_{-0.28}$}
\newcommand{\detwith}{$0.39\pm0.20$}
\newcommand{\bfafwo}{$0.94^{+0.38}_{-0.28}$}
\newcommand{\bfafwith}{$1.17^{+0.30}_{-0.21}$}

\newcommand{\ergs}{\,erg\,s$^{-1}$}

\title[Radio and X-ray observations of \aql]
      {Quasi-simultaneous radio and X-ray observations of \aql\,: probing low luminosities}

 \author[Gusinskaia et al.]
    {N.V. Gusinskaia,$^{1,2}$\thanks{E-mail: N.Gusinskaia@uva.nl}
    J.W.T. Hessels,$^{1,2}$ 
    N. Degenaar,$^{1}$
    A.T. Deller,$^{3}$
 \newauthor J.C.A. Miller-Jones,$^{4}$
    A.M. Archibald,$^{1}$
    C. O. Heinke,$^5$
    J. Mold\'{o}n,$^6$
\newauthor A. Patruno,$^{7,2}$
    J. A. Tomsick,$^8$
    R. Wijnands$^{1}$\\
$^1$Anton Pannekoek Institute for Astronomy, University of Amsterdam, Science Park 904, 1098 XH Amsterdam, The Netherlands\\
$^2$ASTRON, the Netherlands Institute for Radio Astronomy, Postbus 2, 7990 AA, Dwingeloo, The Netherlands\\
$^3$Centre for Astrophysics and Supercomputing, Swinburne University of Technology, P.O. Box 218, Hawthorn, VIC 3122, Australia\\
$^4$International Centre for Radio Astronomy Research, Curtin University. GPO Box U1987, Perth, WA 6845, Australia\\
 $^5$Department of Physics, University of Alberta, CCIS 4-181, Edmonton, AB T6G 2E1, Canada\\
 $^6$Jodrell Bank Centre for Astrophysics, Alan Turing Building, The University of Manchester, Oxford Road, Manchester, M13 9PL, UK\\
 $^7$Institute of Space Sciences (IEEC-CSIC) Campus UAB, Carrer de Can Magrans, s/n 08193 Barcelona, Spain.\\
 $^8$Space Sciences Laboratory, 7 Gauss Way, University of California, Berkeley, CA 94720-7450, USA}

\date{Accepted xxxx xxx xx.  Received xxx xxxx xx; in original form 2016 February 9}

\pubyear{2019}
\begin{document}
\label{firstpage}
\pagerange{\pageref{firstpage}--\pageref{lastpage}}
\maketitle

\begin{abstract}

{\aql} is one of the best-studied neutron star low-mass X-ray binaries. It was previously targeted using quasi-simultaneous radio and X-ray observations during at least 7 different accretion outbursts.  Such observations allow us to probe the interplay between accretion inflow (X-ray) and jet outflow (radio). Thus far, these combined observations have only covered one order of magnitude in radio and X-ray luminosity range; this means that any potential radio---X-ray luminosity correlation, \lr\ $\propto$ \lx$^{\beta}$, is not well constrained ($\beta \approx$ 0.4--0.9, based on various studies) or understood. Here we present quasi-simultaneous Very Large Array and {\it Swift}-XRT observations of {\aql}'s 2016 outburst, with which we probe one order of magnitude fainter in radio and X-ray luminosity compared to previous studies (6$ \times 10^{34} <$ \lx\ $< 3 \times 10^{35}$\,erg\,s$^{-1}$, i.e., the intermediate to low-luminosity regime between outburst peak and quiescence). The resulting radio non-detections indicate that {\aql}'s radio emission decays more rapidly at low X-ray luminosities than previously assumed --- at least during the 2016 outburst. Assuming similar behaviour between outbursts, and combining all available data in the hard X-ray state, this can be modelled as a steep $\beta=$\bfafwith\ power-law index or as a sharp radio cut-off at \lx\ $\lesssim 5 \times 10^{35}$\,erg\,s$^{-1}$ ( given our deep radio upper limits at X-ray luminosities below this value). We discuss these results in the context of other similar studies.
\end{abstract}


\section{Introduction}
\label{sec:intro}

Astrophysical jets (collimated outflows) are an important yet not-well-understood component in the physics of accretion. Despite the development of numerous theoretical models \citep[e.g.,][]{BZ1977,BP1982,Parfrey2016}, their formation mechanism is still unclear. Observations of jets in various types of accreting systems can help to elucidate this problem. A particularly useful class of objects in the study of jet formation are Galactic transient low-mass X-ray binaries \citep[LMXBs;][]{Lewin1997} --- systems consisting of a compact object (which can be either a black hole, BH, or a neutron star, NS), accreting from a low-mass companion star ($\lesssim 1\,\mathrm{M_{\odot}}$ for NS systems, and less then a few solar masses for BH systems). During their characteristic episodes of active accretion towards the central object (outbursts), the accretion rate of LMXBs evolves on time-scales of days to months, allowing us to observe the formation and disappearance of jets on practical timescales.

When a transient LMXB goes into outburst, its multi-wavelength luminosity increases by a few orders of magnitude. As the system's accretion rate increases, accumulated in-falling matter is heated up and emits in X-rays \citep[e.g.,][]{SS1973}. Not all of the in-falling material ends up on the central object, however, some fraction is expelled from the system. In some cases, a jet is formed: outflowing accelerated particles produce synchrotron radiation that can be visible at radio wavelengths. Thus, studying the relationship between the radio luminosity (\lr) and the X-ray luminosity (\lx) of such a system allows one to trace the connection between the inflow and outflow of material in LMXBs --- as well as various other types of accreting systems \citep[e.g., active galactic nuclei,][]{Plotkin2012}. 

Observations of X-ray binaries in the hard X-ray state have shown that \lr\ and \lx\ often follow a power-law relation \lr\ $\propto$ \lx$^{\beta}$ \citep{Corbel2002,MIGFEN2006,Gallo2018}.
This demonstrates that both \lr\ and \lx\ are functions of mass accretion rate, and (under the assumption that the jet power is a fixed fraction of the accretion power, e.g. \citealt{Markoff2003}) the power-law index $\beta$ mostly reflects the properties of the accretion inflow (e.g., how efficient it is at producing radiation; \citealt{Narayan1995}). In such scale-invariant coupled disk-jet models, $\beta\sim 1.4$ is usually associated with radiatively efficient accretion and $\beta\sim 0.7$ with radiatively inefficient accretion. In the case of BH accretors, such radiative inefficiency can be explained by the loss of part of the accreting material within the event horizon \citep{Narayan1995}. In the case of NS accretors, a radiatively inefficient accretion flow could be explained, e.g., by a strong outflow \citep[due to, e.g., propeller effect][]{DAngelo2015}.

It has been argued previously that NS-LMXBs and BH-LMXBs show different relations between their {\lr} and {\lx}, when modelling observations in the hard X-ray state as a power-law relation \citep{MIGFEN2006}.  More recently, however, \citet{Gallo2018} find that NS- and BH-LMXBs show roughly consistent power-law slopes at the $2.5\sigma$ level.  They find $\beta = 0.44^{+0.05}_{-0.04}$ and $\beta = 0.59\pm0.02$ for NS and BH systems, respectively \citep{Gallo2018}.

In comparison to BH-LMXBs, NS-LMXBs are typically radio fainter, by roughly an order-of-magnitude for the same X-ray luminosity \citep{FENKUUL2001,MIGFEN2006,Gallo2018}, and hence more challenging to study.  A particularly interesting NS-LMXB for studying jet activity is {\aql}, one of the first Galactic X-ray sources to be discovered \citep{friedman1967,discovery}. It is a binary system (orbital period $\sim 19$\,hours) composed of a NS \citep{fisrt_typeI_1985} and a low-mass ($\sim 0.4\,\mathrm{M_{\odot}}$) main-sequence-like companion \citep{chevalier1999}. {\aql} goes into outburst almost every year \citep[e.g.,][]{Campana2013,gungor2014}, and hence offers regular opportunities to study its {\lr}---{\lx} relation. 

Based on the X-ray spectral and timing properties, different states are identified during LMXB outbursts \citep[e.g.,][]{VDK1994}.  During its outbursts, \aql\ typically evolves from a low-hard X-ray state, peaking in the high-soft X-ray state, and returning back to the hard state at the end of the outburst \citep{MD2014}.  The majority of outbursts of {\aql} can be characterised by fast rise and exponential decay of the X-ray luminosity (`FRED-type' outburst; \citealt{Campana2013}). Some of {\aql}'s outbursts have symmetrical rise and decay (`Gaussian-like' outburst; \citealt{Campana2013}) or have multiple X-ray luminosity peaks. An alternative classification of outbursts of {\aql} is based on their duration and peak luminosity \citep{gungor2014}: low-luminosity outbursts reach a maximum X-ray luminosity of \lx\ $\sim 10^{37}$\,erg\,s$^{-1}$ before decaying, while high-luminosity outbursts can reach up to a factor of 5 higher in \lx\ and tend to be longer in duration. 

Previous radio monitoring of {\aql} was performed for at least 7 different outbursts (all of the low-luminosity class). These studies demonstrated that its radio jet behaviour changes depending on the X-ray spectral state and luminosity \citep{MJ2010,TUD2009}. Flat-spectrum\footnote{ Assuming a power-law spectrum, where radio flux density, $S_{\nu}$, is proportional to frequency, $\nu$, as $S_{\nu} \propto \nu^{\alpha}$, and where $\alpha \approx 0$ is defined as flat.} radio emission (indicative of an optically thick jet) was observed during the rise in the low-hard X-ray state or transition between low-hard and high-soft X-ray states \citep{MJ2010,diaztrigo2018}, and `quenches' (i.e. disappears) during the peak of the high-soft X-ray state (\lx\ $\sim10^{37}$\,erg\,s$^{-1}$; \citealt{MJ2010}).

{\aql} is one of only 3 NS-LMXBs in which the correlation between radio and X-ray luminosities has been investigated in detail using a relatively large number of observations. The power-law index was fitted in different studies, using different data selections (see \S\ref{sec:sel_fit}), which found $\beta=0.40\pm 0.07$ \citep{TUD2009} and $\beta=0.76^{+0.14}_{-0.15}$ \citep{Tetarenko2016}.  This is similar to what has been found in NS- and BH-LMXBs when fitting the populations globally and accepting more scatter in the correlation \citep{Gallo2018}. For two other NS-LMXBs, 4U~1728$-$34 \citep{MIG2003} and EXO~1745$-$248 \citep{Tetarenko2016}, steeper power-law indices were found: $\beta=1.5\pm0.2$ and $\beta=1.68^{+0.10}_{-0.09}$, respectively.  This could suggest a radiatively efficient accretion regime \citep{MIGFEN2006}.  Different power-law indices have also been observed in individual BH-LMXBs (e.g., the `outliers track' described by \citealt{Coriat2011}); however, they have been observed to return to the same canonical track for \lx\ $< 10^{35}$\,erg\,s$^{-1}$ \citep{Corbel2013,Maccaone2012}.

The situation can be clarified by observing such systems over the widest-possible range of radio and X-ray luminosities.  While the radio---X-ray correlation was probed for a few BH-LMXBs over 8\,dex (orders of magnitude) in \lx\ and 5\,dex in \lr, in 4U~1728$-$34 and {\aql} this correlation has previously only been studied over $\sim1$\,dex in radio and X-ray luminosities.

Since NS-LMXBs tend to be more radio faint compared to BH-LMXBs \citep{FENKUUL2001,Gallo2018}, radio observations below the peak of their outburst (\lx\ $<10^{36}$\,erg\,s$^{-1}$) have previously been hampered by the limited sensitivity of available radio telescopes. With the upgraded Karl J. Jansky Very Large Array \citep[VLA;][]{Perley2011}, an order-of-magnitude deeper radio observations are now possible. However, such observations are still challenging to schedule because NS-LMXBs often pass through the $10^{34-36}$\,erg\,s$^{-1}$ X-ray luminosity range only during the rise and decay of the outburst, which often lasts no longer than $\sim 3-7$ days \citep{Campana2013,Gungor2017}.

Due to these time constraints and their relative radio faintness, only a few NS-LMXBs have been observed in the luminosity range \lx\ < $10^{36}$ erg\,s$^{-1}$. In the case of non-pulsating NS-LMXBs, these observations led exclusively to non-detections \citep{gusinskaia2017,Tetarenko2016,Tudor2017}. However, there are two classes that potentially stand out in this low X-ray luminosity regime \citep{Gallo2018}: accreting millisecond X-ray pulsars (AMXPs; systems that exhibit coherent millisecond X-ray pulsations as a result of accreting material channelling onto the magnetic polar caps of the NS; \citealt{WIjnands1998,Patruno2012}) and transitional millisecond pulsars (tMSPs; NSs that switch between rotation-powered radio millisecond pulsars and an accretion-disc state; \citealt{Papitto2013,Jaodand2018}). Both these classes show comparatively bright radio emission at low \lx\ \citep{DEL2015,Tudor2017}. There were previously tentative suggestions that tMSPs might follow their own correlation in the radio---X-ray luminosity plane \citep{DEL2015}. However, recent strictly simultaneous radio and X-ray observations of tMSP PSR~J1023+0038 in the low-luminosity accretion-disc state showed anti-correlated radio---X-ray brightness variations that argue against an origin from a collimated jet \citep{Bogdanov2018}.  Thus, the accretion-disc state of low-luminosity tMSPs is likely quite different compared to NS-LMXBs in outburst. Consequently, care is needed when comparing different NS-LMXB systems in the \lr---\lx\ plane because they could be in fundamentally different accretion regimes.

Ultimately, many more systems need to be studied in detail before we can come to any robust conclusions on whether different classes of NS-LMXBs have different \lr---\lx\ correlations, because of, e.g., the influence that different magnetic field strengths, geometry or spin rate may have on disk-jet coupling.

\begin{figure*}
\centering
\includegraphics[width=\textwidth]{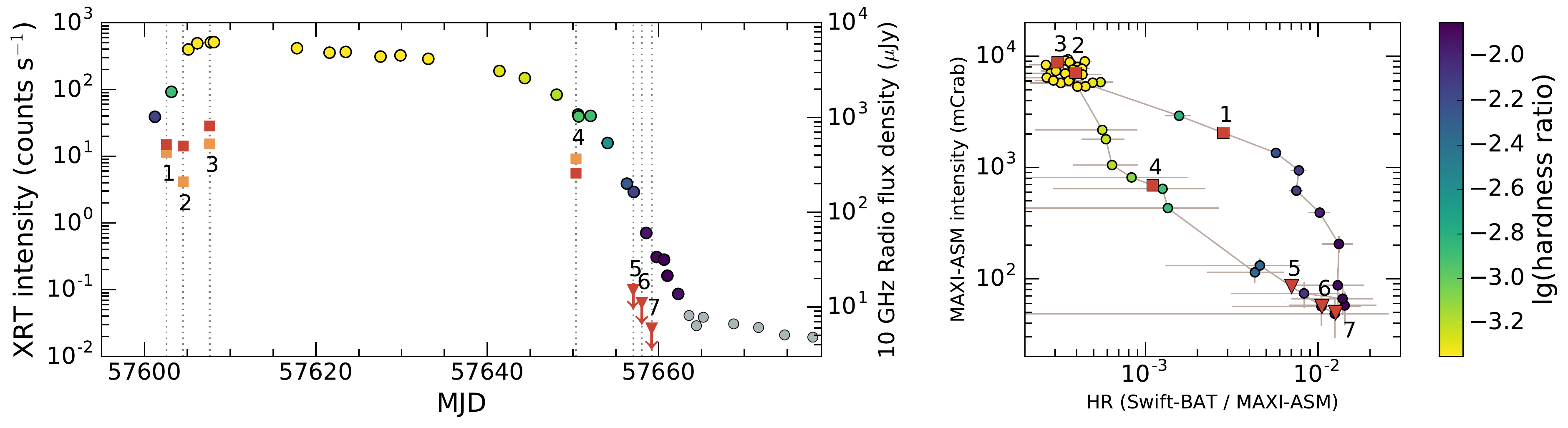}

\caption{\textbf{\textit{Left:}} X-ray and radio light-curves of {\aql}'s 2016 outburst. Circular symbols represent the {\it Swift}-XRT ($0.3-10$\,keV) X-ray light-curve (using the left-hand axis). Their colour represents hardness ratio (see colourbar on the right-hand side of this figure; grey symbols indicate no constraint on the spectrum). Square symbols represent the ATCA \citep{diaztrigo2018} radio light-curve (using the right-hand axis): red and orange symbols are 5.5 GHz and 9.0 GHz flux densities, respectively. Red downward-pointing triangles are 10-GHz VLA upper limits from our campaign. Numbers refer to radio observations in Table~\ref{tab:obs}, second column. The dotted vertical lines indicate the epochs of the radio observations and the associations with their nearest corresponding X-ray observations. \textbf{\textit{Right:}} Hardness-intensity diagram of the same outburst (circular symbols). The hardness ratio is defined as $15-50$\,keV {\it Swift}-BAT flux / {\it MAXI}~$2-10$\,keV flux and intensity as $2-10$\,keV {\it MAXI} flux in mCrab units. The red symbols indicate at what stage of the outburst the radio observations were taken (squares: ATCA and triangles: VLA), and use the same numbering convention as in the left panel and Table~\ref{tab:obs}.}\label{fig:aql_lc}
\end{figure*}

\aql\ is identified as an intermittent AMXP because of the discovery of $550.27$\,Hz coherent X-ray pulsations \citep{CAS2008} that lasted for only 150 seconds out of the $10^6$ seconds that the source was observed. Two other AMXPs are also classified as intermittent and have shown pulsations that appear and disappear on time-scales of days to years: HETE~J1900.1$-$2455 \citep{Galloway2007,Patruno_hete2012} and SAX~J1748.9$-$2021 \citep{Altamirano2008}.  Since \aql\ is the most extreme case of intermittency, it is also possible that the origin of the pulsations is different in nature.

In this paper we present sensitive radio observations of \aql\ in the low X-ray luminosity range ($< 10^{36}$\,erg\,s$^{-1}$), taken while the source was decaying from its 2016 outburst.  We compare these with archival measurements from this and previous outbursts to study the \lr---\lx\ correlation of \aql\ --- under the assumption that the radio---X-ray correlation behaves similarly between outbursts.  In \S\ref{sec:obs} we present the observations and data analysis.  In \S\ref{sec:results} we present the results, and discuss their interpretation in \S\ref{sec:disc}.

\vspace{-7mm}

\section{Observations and data reduction}
\label{sec:obs}

\begin{table*}
\caption[All observations]{VLA radio upper
  limits (3$\sigma$), ATCA detections \citep{diaztrigo2018}, together with the corresponding
  quasi-simultaneous {\it Swift}-XRT X-ray observations and their spectral properties. All uncertainties are 1-$\sigma$.}

\begin{minipage}{177mm}
{
\renewcommand{\arraystretch}{1.7}
\begin{tabular}{@{\extracolsep{-2pt}}l|cccc|cccccc@{}}
\hline\hline
\multicolumn{1}{c}{} & \multicolumn{4}{c}{ATCA Radio} & \multicolumn{6}{c}{{\it Swift}-XRT X-ray ($1-10$ keV)} \\
\hline
   \makecell[c]{Date\\ 2016}  & \#&\makecell[c]{ MJD } & \makecell[c]{$S_{\nu}$ \\ ($\upmu$Jy)} &  \makecell[c]{$\nu$\\(GHz)} & \makecell[c]{MJD } & \makecell[c]{Obs. ID and \\ mode} & \makecell[c]{Spectral\\ model$^a$} & \makecell[c]{Unabsorbed \\ flux  $\times 10^{-10}$\\($\mathrm{erg\,s^{-1}\,cm^{-2}}$)}   &  \makecell[c]{Photon \\ index} &  $\chi^2_{\nu}$ (dof)  \\
\hline
 2 Aug & 1 &57602.543 & \makecell[c]{428 $\pm$ 28 \\ 512 $\pm$ 26} &  \makecell[c]{9\\5.5}& ---  & --- & --- & --- & --- & --- \\
 3 Aug & --- & --- & --- & --- & 57603.137  &  33665074(wt)& \texttt{TBabs*(po)} & $60\pm1$    & $1.62\pm0.02$ & 1.01 (473)\\
 4 Aug & 2 & 57604.509 & \makecell[c]{208 $\pm$ 18\\ 498 $\pm$ 19} & \makecell[c]{9\\5.5}& --- & --- & --- & --- & --- & --- \\
 5 Aug & --- & --- & --- & --- & 57605.111  &  33665075(wt)& \texttt{TBabs*(po+bb)} & $246\pm2$ & $1.65\pm0.02$ & 1.06 (674)\\
 6 Aug & --- & --- & --- & --- & 57606.140  &  33665078(wt)& \texttt{TBabs*(po+bb)} & $357\pm3$    & $1.56\pm0.02$ & 1.02 (676)\\
 7 Aug & 3 &57607.599 & \makecell[c]{528 $\pm$ 19\\ 810 $\pm$ 19} & \makecell[c]{9\\5.5}&57607.726  &  33665076(wt)& \texttt{TBabs*(po+bb)} &$313\pm5$    & $1.81\pm0.05$ & 1.01 (500)\\
19 Sep & 4 &57650.354 & \makecell[c]{366 $\pm$ 12 \\ 259 $\pm$ 13} &  \makecell[c]{9\\5.5}& 57650.597&  33665089(wt)& \texttt{TBabs*(po+bb)} & $18.7\pm0.3$    & $1.95\pm0.05$ & 1.03 (397)\\
\hline
\multicolumn{1}{c}{} & \multicolumn{4}{c}{VLA Radio} & \multicolumn{6}{c}{{\it Swift}-XRT X-ray ($1-10$ keV)} \\
\hline
25 Sep & --- & --- & --- & --- & 57656.293  &  34719004(pc) & \texttt{TBabs*(po)} &$1.9\pm0.1$    & $1.77\pm0.09$ & 1.11 (50)\\
26 Sep & 5  &57657.031 & <15&10& 57657.095  &  34719005(pc)& \texttt{TBabs*(po)} & $1.2\pm0.1$    & $1.87\pm0.12$ & 1.08 (30)\\
27 Sep & 6 & 57658.031 & <11&10& 57658.558  &  34719006(pc)& \texttt{TBabs*(po)} & $0.36\pm0.04$    & $1.93\pm0.16$ & 1.35 (16)\\
28 Sep & 7 &57659.181 & <6 &10& 57659.767  &  34719007(pc)& \texttt{TBabs*(po)} & $0.16_{-0.05}^{+0.08}$    & $2.33^{+0.21}_{-0.19}$ & 1.25 (10)\\
\hline
\end{tabular}
}
\begin{flushleft}{
$^{a}$ We used $N_{\rm H} = 0.4 \times 10^{22}$\,cm$^{-2}$ as a fixed parameter.
  }\end{flushleft}
\end{minipage}
\label{tab:obs}
\end{table*}

\subsection{2016 outburst}

\aql\ entered into outburst at the end of July 2016 \citep{ATel_outb_start}. The outburst lasted for approximately two months, reaching a peak X-ray luminosity of $\sim 8 \times  10^{37}$\ergs\ (for an assumed distance of 4.5 kpc; \citealt{Gungor2017}), making it the brightest outburst observed from \aql\ to date.

Using the VLA, we performed three radio observations of \aql\ while it was fading in brightness at the end of its 2016 outburst. These observations were spaced daily (on the 26\thh, 27\thh\ and 28\thh\ of September, 2016), and were coordinated quasi-simultaneously with Neil Gehrels {\it Swift}-XRT daily monitoring. Table~\ref{tab:obs} provides a log of all 2016 outburst observations used in the present study.

\subsubsection{Radio data and analysis}

\aql\ was observed with the VLA at 3 epochs (project ID: 15B-239), each at X-band ($8-12$\,GHz). The first two epochs had total duration $\sim 1$\,h and the last was $\sim 3$\,h including calibration scans, with $\sim 0.5$\, and $\sim 2$\,hours on source, respectively. During all three epochs the array was in A configuration (synthesised beam $\sim 0.3$\arcsec). Data were initially calibrated using the VLA Common Astronomy Software Application (CASA\footnote{\url{https://casa.nrao.edu/}}; \citealt{McMullin2007}) Calibration Pipeline (version 5.4.1).  We used J1331+305 and J1407+2807 as flux and polarisation calibrators, respectively, and J1907+0127 as phase and amplitude calibrator. The data were additionally flagged and calibrated following standard procedures within CASA.

In our analysis below, we also incorporate results from Australia Telescope Compact Array (ATCA) observations of \aql's 2016 outburst that were previously presented in \citet{diaztrigo2018}.

\subsubsection{X-ray data and analysis}
\label{xray_analysis}

Based on the 3 observational epochs obtained using the VLA we selected 4 quasi-simultaneous {\it Swift}-XRT observations (target ID: 34719), all done in Photon Counting (PC) mode.  As such, each of our radio observations is framed by a pair of X-ray observations. Additionally, we selected 7 {\it Swift}-XRT observations (target ID: 33665) --- 2 done in PC mode and 5 done in Window Timing (WT) mode --- that are spaced quasi-simultaneously between the aforementioned ATCA radio observations \citep{diaztrigo2018}.  For more details see Table~\ref{tab:obs} and Figure~\ref{fig:aql_lc}.  Due to the high count rate of \aql, all observations were corrupted by pile-up, and consequently 2 observations that were performed in PC mode (observation IDs: 33665073 and 33665088) during the peak of the outburst were excluded.

The basic data reduction and calibration was done using the {\tt XRT$\_$PIPELINE} task within {\tt HEASoft-6.24}\footnote{\url{https://heasarc.gsfc.nasa.gov/docs/software/heasoft/}}. For all 9 epochs, the spectrum of the source was extracted from a 30-pixel-wide area (circular in the case of PC mode observations and square in the case of WT mode observations), centred on the source position (taken from \citealt{MJ2010}), while the background was extracted from a surrounding annulus with 30-pixel inner radius and 60-pixel outer radius. The extraction of spectra was done using {\tt XSELECT}. To correct for known CCD artefacts we used ancillary arf-files (produced using the {\tt XRTMKARF} script and exposure map) together with the
`swxpc0to12s6$\_$20130101v014.rmf' response file. Each spectrum was grouped into 20-photon bins. In order to account for pile-up, we used the technique described in \citet{pile-up} and excluded the necessary amount of pixels from the central part of the images (or events in the case of WT mode) before performing spectral model fitting. We used {\tt XSpec} (version 12.10.0c) to perform the spectral model fit and to extract the source flux. For spectral fitting we used photons from the $0.4-10$\,keV and  $0.7-10$\,keV ranges in the case of PC and WT mode observations, respectively. We achieved sufficiently good fits ($\chi^2_{\nu} < 1.1$) using a simple absorbed power-law model ({\tt TBabs * powerlaw}) or with an additional black body component ({\tt TBabs * (powerlaw + bb)}) in {\tt XSpec}. By fitting all spectra together with the $N_{\rm H}$ parameter tied between data sets, we found $N_{\rm H} = 0.39\pm 0.02 \times 10^{22}$\,cm$^{-2}$, which is consistent with the value reported by \citet{Rutledge2001}. Thus we simply used $N_{\rm H} = 0.4 \times 10^{22}$\,cm$^{-2}$ as a fixed parameter while performing fits to individual observations. We added a {\tt cflux} component to the spectral model in order to determine unabsorbed $1-10$\,keV fluxes and their associated errors for each epoch. The resulting parameter values, spectral models, $\chi^2_{\nu}$ values and number of degrees of freedom (dof) are listed in Table~\ref{tab:obs}.

In order to place our radio observations in context, we produced a hardness-intensity diagram (HID) using daily-averaged count rates of all-sky monitors {\it MAXI} \footnote{\url{http://maxi.riken.jp/mxondem/20190423121006_5c4blcsw66n1gu78/index.html}} 
and {\it Swift}-BAT\footnote{\url{https://swift.gsfc.nasa.gov/results/transients/AqlX-1/}}. Both count rates were converted to mCrab using the average (over the time period of \aql's 2016 outburst) count rate of the Crab nebula with the corresponding instrument. Here we define hardness ratio as $15-50$\,keV {\it Swift}-BAT flux / {\it MAXI} $2-10$\,keV flux and intensity as $2-10$\,keV {\it MAXI} flux. The obtained HID is shown in Figure~\ref{fig:aql_lc} (right). The colour scale represents the hardness ratio, where yellow is more soft and blue is more hard. For illustration purposes we inferred the hardness ratio for each {\it Swift}-XRT observation by interpolating the results from {\it MAXI}/{\it Swift}-BAT observations taken around the time of the {\it Swift}-XRT observations.

In order to directly compare the 2016 outburst with the other 7 outbursts (which were observed using {\it RXTE}-PCA) we additionally produced HIDs where the intensity and the hardness ratio are defined as $2-16$\,keV flux and ratio of $9.7-16$\,keV and $6.0-9.7$\,keV fluxes, respectively. Analogous to the above, we converted {\it MAXI} count rates into mCrab units using the average (during the 2016 outburst) count rate of the Crab nebula in the corresponding energy bands. The mCrab units were then converted to $\mathrm{erg\,s^{-1}\,cm^{-2}}$ flux (see Figure~\ref{fig:all_hids}).

As shown in Figure~\ref{fig:aql_lc}, {\aql} was in the hard X-ray state during all VLA observations. The first VLA observation was performed just 6 days after the last ATCA detection, which was performed a few days after {\aql} had started the spectral transition to the hard state. As such, two ATCA observations of {\aql}'s 2016 outburst were performed when the source was still relatively hard and two when it was in the soft X-ray state.

\subsection{Archival data of \aql\, from previous outbursts}
\label{subsec:archive}

\subsubsection{Radio data}
\label{archive:radio}

In our analysis below (\S\ref{sec:results}), we combine data from the 2016 outburst with previous radio---X-ray monitoring campaigns of \aql. We used the results of all published radio continuum observations that were taken before 2011 (Table~1 from \citealt{TUD2009} and the entire\footnote{\url{https://iopscience.iop.org/2041-8205/716/2/L109/suppdata/apjl346296t1_mrt.txt}} Table~1 from \citealt{MJ2010}). From this set we selected only radio observations that were taken within one day of an {\it RXTE}-PCA X-ray observation.

\vspace{-3mm}

\subsubsection{X-ray data}
\label{archive:xray}

From the {\it RXTE}-PCA data archive we selected the closest observation to each selected radio point. In most cases it was possible to bracket the radio observation with an additional X-ray observation that was within two days of the radio measurement, and which allowed us to interpolate the X-ray flux (see \S\ref{sec:res_lums}).

Selected X-ray observations were analysed using the standard {\tt HEASoft} tools. Each spectrum was produced by grouping 20 photons per bin, subtracting background, applying telescope response corrections and --- in the case of 7 observations --- removing Type-I X-ray bursts. In order to correct for the systematic uncertainty of the {\it RXTE}-PCA \citep{rxte_syst} energy bins, we increased the uncertainties of each spectral bin by 0.5\% (following e.g. \citealt{Nowak_syst}). We performed basic spectral fitting with {\tt XSpec} using three different possible models: a power-law, a power-law plus a Gaussian (if the iron K-$\alpha$ line was present), and a power-law plus a black-body component (if an additional thermal component was needed to produce an acceptable fit). For this spectral fitting we only used the $3-16$\,keV band; thus we kept the $N_{\rm H}$ value fixed to the value quoted above. For the neutral iron line modelling we restricted the mean of the Gaussian to the range $6.3-6.5$\,keV (since the iron line appears at 6.4\,keV). We achieved acceptable fits for all observations ($\chi^2_{\nu} < 1.3$ and p-value $>0.08$). We then used the {\tt cflux} model to obtain {\aql}'s unabsorbed $1-10$\,keV flux. We also obtained the unabsorbed $2-16$\,keV, $6-9.7$\,keV and $9.7-16$\,keV fluxes from each spectrum in order to calculate the intensity and hardness ratio (HR) in the same way that it was defined in \citet{MJ2010}. The resulting fluxes and hardness ratios for each observation are presented in Table~\ref{tab:appendix}. We repeated this analysis for all {\it RXTE}-PCA observations in all 7 previous outbursts discussed in this paper. The associated light-curves and HIDs for all previous outbursts are presented in the Appendix (Figures~A1---A7). 

\begin{figure*}
\centering
\includegraphics[width=\textwidth]{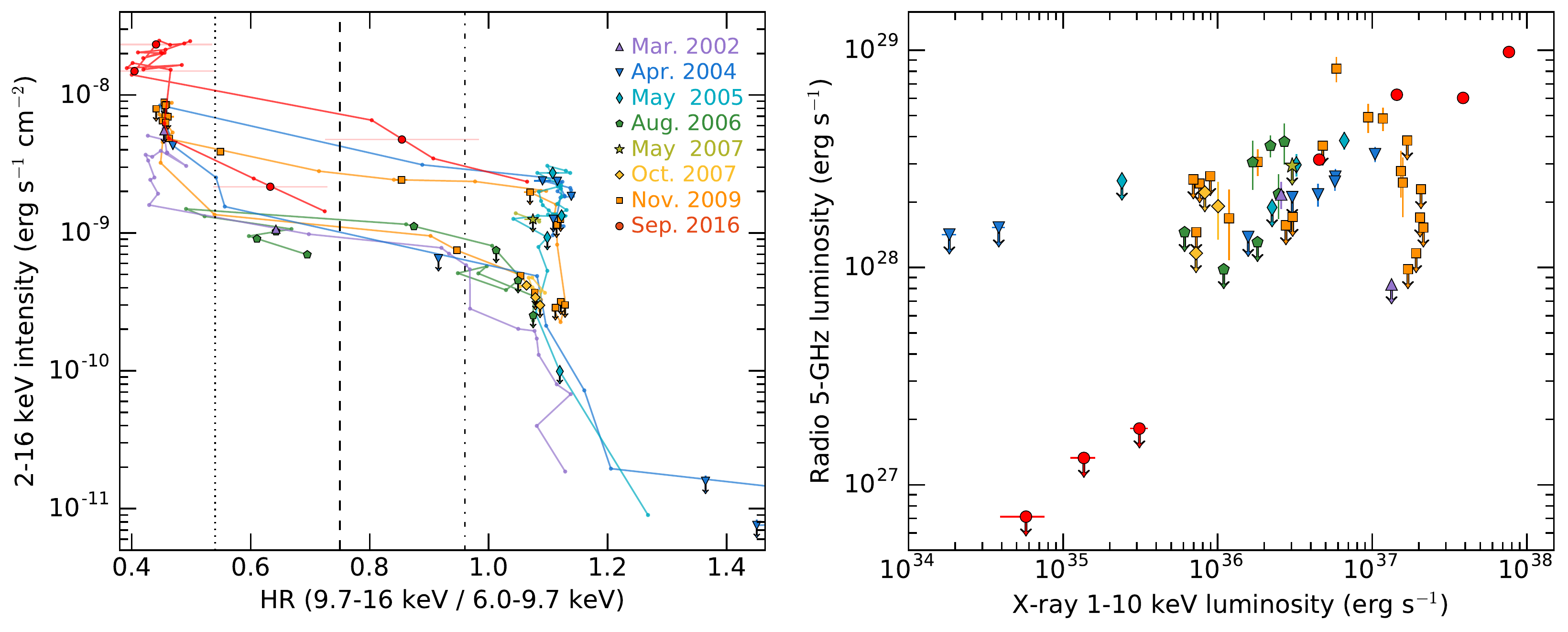}

\caption{\textbf{\textit{Left:}} Hardness-intensity diagram for the 8 outbursts of Aql X-1 used in this study. Different symbols and colours indicate X-ray properties at the time of radio observations from different outbursts. The hardness ratio for outbursts before 2016 is defined from {\it RXTE}-PCA data. For the 2016 outburst, the hardness ratio was defined from {\it MAXI} data (note that, for flux less than $10^{-9} \mathrm{erg\,s^{-1}\,cm^{-2}}$, the {\it MAXI} count rate in the $9.7-16$\,keV and $6.0-9.7$\,keV energy bands was too low to constrain the spectrum but, from the joint {\it MAXI} and {\it Swift}-BAT HID, we see that the VLA upper-limits are during the hard X-ray sate, i.e. HR $\gtrsim 1$). Vertical lines represent the hardness ratio threshold that we used in our data selection: dotted line HR = 0.54, dashed line HR = 0.75 and dash-dotted line HR = 0.96. \textbf{\textit{Right:}} Quasi-simultaneous X-ray ($1-10$\,keV) luminosity versus radio (5\,GHz) luminosity for Aql X-1. Different symbols and colours indicate different outbursts, using the conventions as in the left panel.  Radio data points are taken from \citet{TUD2009} and \citet{MJ2010} for outbursts before 2016, and from \citet{diaztrigo2018} for radio detections during the 2016 outburst.}
\label{fig:all_hids}
\end{figure*}

\section{Analysis and results} 
\label{sec:results}

In the following analysis, we assume a distance of 4.5\,kpc --- as was done in \citet{Campana2014} based on the results of \citet{Galloway2008}, where the peak luminosity of Type-I X-ray bursts was used to estimate distance.

\subsection{Radio and X-ray luminosities}
\label{sec:res_lums}
All three of our VLA observations of {\aql} resulted in non-detections.  Using the {\tt imfit} task in CASA and the exact position of the source (taken from \citealt{MJ2010}), we measured flux densities of $2.5 \pm 4.2$\,$\upmu$Jy/beam, $-1.9 \pm 4.3$\,$\upmu$Jy/beam and $-0.2 \pm 2.0$\,$\upmu$Jy/beam at the three epochs, respectively, where the error corresponds to the off-source local rms image noise.  These correspond to 3-$\sigma$ upper limits of 15\,$\upmu$Jy/beam, 11\,$\upmu$Jy/beam and 6\,$\upmu$Jy/beam, respectively (Table~\ref{tab:obs}).

{\aql} was detected in all {\it Swift}-XRT observations that are used in our study (Table~\ref{tab:obs}). During the 2016 outburst, {\aql} had an X-ray luminosity typical of NS-LMXBs in outburst ($10^{36-38}$\,erg\,s$^{-1}$). Nonetheless, we note that the 2016 outburst was a particularly high-luminosity outburst --- in fact the brightest observed from \aql\ to date --- reaching a peak X-ray luminosity of almost $10^{38}$\,erg\,s$^{-1}$ \citep{Gungor2017}. Our VLA observations were performed as the source was decaying in brightness, with the first VLA observation taken when {\aql} was at an X-ray luminosity of $3 \times 10^{35}$\,erg\,s$^{-1}$ and the last one when it was at $6 \times 10^{34}$\,erg\,s$^{-1}$.

Since radio and X-ray observations were not strictly simultaneous, we obtained the X-ray flux values corresponding to the 2016 VLA and ATCA radio observations by logarithmically interpolating between pairs of closely spaced {\it Swift}-XRT observations (see Figure~\ref{fig:aql_lc}). In the same manner, we obtained approximate X-ray fluxes and hardness ratios ($9.7-16$\,keV flux/$6-9.7$\,keV flux) at the time of each archival radio observation (\S\ref{archive:radio}) using {\it RXTE}-PCA data (\S\ref{archive:xray}).  We increased the uncertainties on the interpolated X-ray fluxes at the times of the radio epochs in order to account for potential short-timescale ($< 1$\,day) X-ray variability.  Using {\it RXTE}-PCA data from all 7 archival outbursts of \aql\, (190 observations), we found that the average maximum fractional difference in fluxes of two adjacent X-ray observations ($\Delta_{fl}=|fl_1 - fl_2|/fl_1$) is proportional to the time between observations ($\Delta_{t}$ in days) in the form: $\Delta_{fl}=0.2 \times \Delta_t +0.005$.  We therefore increased the uncertainty on each interpolated X-ray flux (typically by $\sim 6$ times compared to the formal uncertainties) using this formula and the separation in time between the radio epoch and closest available X-ray observation.  When the source was dim, the formal errors on the X-ray fluxes dominated.

The 1--10\,keV X-ray fluxes were converted into luminosities using  $ L_\mathrm{X} = 4 \pi D^2 S$, where $S$ is the observed flux and $D$ is distance to the source. The 5\,GHz radio luminosities were calculated using $ L_\mathrm{R} = 4 \pi \nu D^{2} S_\nu $, where $\nu$ is the central frequency and $S_\nu$ is the observed flux density. To incorporate radio measurements at different frequencies ($\sim 4-12$\,GHz) we assumed a flat spectral index, which is consistent with previous multi-band radio measurements of \aql\,in the hard X-ray state \citep{MJ2010,diaztrigo2018}.

A summary of all 2016 and archival radio---X-ray measurements used in this study is provided in Table~\ref{tab:appendix}.  The figures of the Appendix (Figures~A1---A7) show lightcurves, hardness-intensity diagrams and \lr---\lx\ diagrams for each of the archival outbursts individually.

\begin{figure*}
\centering 
\includegraphics[width=\textwidth]{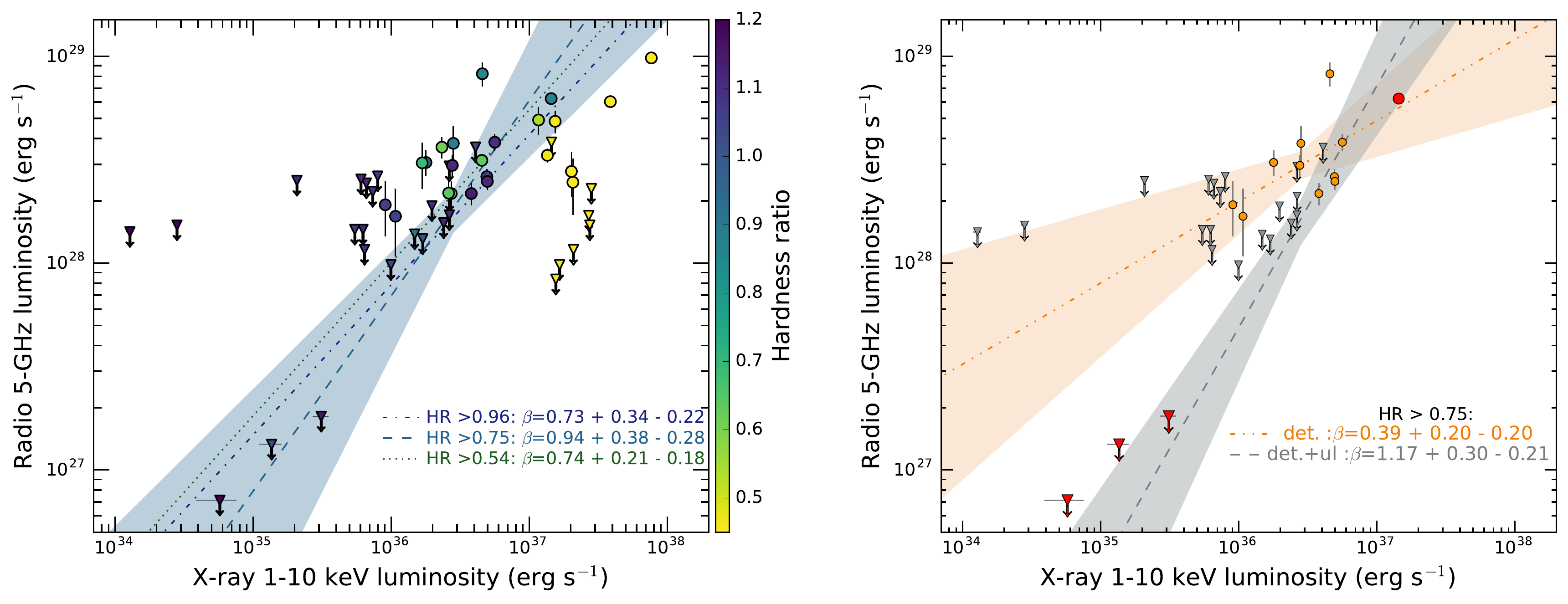} 

\caption{Quasi-simultaneous X-ray ($1-10$\,keV) luminosity versus radio (5\,GHz) luminosity for Aql X-1. Circles and downward-pointing triangles represent radio detections and upper-limits, respectively. \textbf{\textit{Left:}} All radio---X-ray observations used in our study. Their colours represent HR, defined as the ratio of  $9.7-16$\,keV and $6.0-9.7$\,keV fluxes. Different lines represent the results of the fit using different HR thresholds and using only the archival data (i.e., the 2016 outburst data are not used in the fits shown here, but the data points are shown nonetheless in order to demonstrate their consistency with the fit to archival data points): dotted line HR $> 0.54$, dashed line HR $> 0.75$ and dash-dotted line HR $> 0.96$.  The blue-shaded area is the inferred 1-$\sigma$ uncertainty on the resulting slope for HR $> 0.75$. \textbf{\textit{Right:}} All radio---X-ray observations with HR $> 0.75$ are included in the fits shown here. Orange and grey symbols indicate archival detections and non-detections, respectively. Red symbols indicate the data from the 2016 outburst. The power-law fit using only radio detections is shown by the dash-dot-dotted line, with the orange shaded area representing its 1-$\sigma$ uncertainty. Similarly, the dashed grey line and grey shaded area represents the power-law fit to the data using both radio detections and upper-limits. Radio data points are taken from \citet{TUD2009}, \citet{MJ2010} for outbursts before 2016 and from \citet{diaztrigo2018} for radio detections during the 2016 outburst.}\label{fig:lrlx_new}

\end{figure*}

\subsection{Radio---X-ray correlation}

\subsubsection{Fitting method}

In order to investigate the radio---X-ray correlation we performed a power-law fit to the luminosities in the form used by \citet{Gallo2014}:
\begin{equation}
\frac{ L_\mathrm{R} }{ L_\mathrm{R,c} } = \xi \left(\frac{ L_\mathrm{X} }{ L_\mathrm{X,c} } \right)^\beta,
\end{equation}
where $L_\mathrm{R,c} = 2.95 \times 10^{28}$ and $L_\mathrm{X,c}=2.82 \times 10^{36}$ (\ergs) are geometrical averages of the radio and X-ray luminosities (calculated using only detections), $L_\mathrm{R}$ and $L_\mathrm{X}$ are radio and X-ray luminosities of all observations (including upper limits), $\beta$ is the power-law index,
and $\xi$ denotes an arbitrary scaling factor.

Following \citet{Gallo2014}, we used the {\tt LINMIX} method developed by \citet{linmix}, which also allows the inclusion of upper limits in the fit \citep[see \S5.2 of][]{linmix}. This method performs a Bayesian-based Markov Chain Monte Carlo (MC-MC) fit of a linear model to the data in logarithmic space:
\begin{equation}
\lg\,L_\mathrm{R}-\lg\,L_\mathrm{R,c} = \lg\,\xi + \beta \left(\lg\,L_\mathrm{X}-\lg\,L_\mathrm{X,c}\right)
\end{equation}
We used the {\tt Python} implementation of the \citet{linmix} linear regression algorithm {\tt LINMIX\_ERR}\footnote{\url{https://github.com/jmeyers314/linmix}}. This tool takes logarithmic values of {\lr} and {\lx} with their uncertainties (or upper-limit values of \lr\ in the case of non-detections) and performs a fit for three free parameters: $\beta$, $\xi$ and an additional parameter $\sigma_{0}$ that accounts for an intrinsic random (Gaussian) scatter of the luminosity values around the best-fit power law.  

As in \citet{Gallo2018}, we calculated the median values of $\beta$, $\xi$ and $\sigma_{0}$ from 10,000 draws of the posterior distribution, and determined $1\sigma$ confidence intervals using the $16-84$th percentiles of the posterior distributions.  Additionally, we performed 500 realisations of this method in order to confirm that the results were robust.  The quoted values in Table~\ref{tab:fit_res} are the average of these realisations.

\subsubsection{Selection criteria and fit results}
\label{sec:sel_fit}
Differences in the criteria used to select what data points are included in the fitting procedure can lead to significant differences in the resulting inferred power-law index. Such selection effects are, in fact, responsible for the difference in power law indices previously reported in the literature: $\beta = 0.40\pm0.07$ \citep{TUD2009} and $\beta = 0.76^{+0.14}_{-0.15}$ \citep{Tetarenko2016}. \citet{TUD2009} used radio---X-ray observations of {\aql} performed during its 2002, 2004 and 2005 outbursts, including all radio observations, regardless of X-ray state. \citet{Tetarenko2016} used data points from the 2004, 2006, 2007 and 2009 outbursts of {\aql}, but excluded radio observations that were performed in the soft X-ray state (HR $< 0.75$ according to their definition, which is the same as used here). In both cases the authors did not include radio upper-limits when performing their fits.

For our analysis, we used data from all 8 outbursts for which quasi-simultaneous radio---X-ray data are available (see Figure~\ref{fig:all_hids}, Table~\ref{tab:appendix}, and Figures~A1---A7).  According to the classification of \citet{Campana2014}, the 2002, 2004, 2005, 2006 and 2016 outbursts are FRED-like; the May 2007 outburst is multi-peaked; and the October 2007 and 2009 outbursts are Gaussian-like.  Alternatively, \citet{Gungor2017} classifies all these outbursts as low-luminosity, with the exception of the 2016 outburst.  As in previous studies, we assume that --- regardless of the outburst classification --- the \lr---\lx\ correlation behaves similarly between outbursts.  We do this because of the relatively low number of available observations, and argue that this is justified by the fact that these measurements occupy the same region in the \lr---\lx\ diagram (within the scatter), without obvious systematic offsets between outbursts (Figure~\ref{fig:all_hids}).

We performed fits using three different HR thresholds (HR $> 0.54$, 0.75 and 0.96), both including and excluding upper-limits, as well as with and without the 2016 outburst --- see Table~\ref{tab:fit_res} for a summary.  
The data points corresponding to the various selection criteria are shown in Figure~\ref{fig:lrlx_new}.

The inferred slope changed with HR threshold, but within the uncertainties (Fig.~\ref{fig:lrlx_new} and Table~\ref{tab:fit_res}). To allow direct comparison with previous studies, we adopt the threshold value of HR $> 0.75$.  The inclusion of upper limits produces a steeper slope compared to that previously reported in the literature, even if the 2016 outburst is left out.  The inclusion of the 2016 outburst also leads to a consistent steep power-law index, that is consistent with the strong VLA radio upper limits at relatively low X-ray luminosity.  In summary, our preferred fit uses HR $> 0.75$, and includes upper limits as well as the 2016 outburst data.

\begin{table*}
\caption[All observations]{Results of the power-law fit for different selections of observations.
The first column indicates the assumed HR threshold for data selection. The sixth column indicates whether the 2016 outburst
data were included (``+ sign) or excluded (``- sign) while performing the fit.
All uncertainties are 1-$\sigma$. Highlighted in bold is our preferred fit, which uses HR $> 0.75$, and includes upper limits as well as the 2016 outburst data.} \label{fit_res}
\begin{minipage}{180mm}
\begin{center}
{
\renewcommand{\arraystretch}{1.5}
\begin{tabular}{@{\extracolsep{2pt}}c|ccc|ccc|c@{}}
\hline\hline
\multicolumn{1}{c}{} & \multicolumn{3}{c}{Only detections} & \multicolumn{3}{c}{Detections + upper-limits} \\ 
\hline
 \makecell[c]{HR \\ $>$} & \makecell[c]{Index\\  $\beta$} &  \makecell[c]{Scatter \\ $\sigma_{0}$ (dex)} &
\makecell[c]{Intercept\\  $\xi$ }&
\makecell[c]{Index\\  $\beta$}  & \makecell[c]{Scatter \\ $\sigma_{0}$ (dex)}  &  \makecell[c]{Intercept\\  $\xi$} &
\makecell[c]{2016\\ outburst} \\ 
\hline
\multirow{2}{*}{\makecell[c]{0.54}} &$0.32\pm0.18$&$0.03^{+0.02}_{-0.01}$&$0.00\pm0.05$&
$0.74^{+0.21}_{-0.18}$&$0.05^{+0.04}_{-0.02}$&$-0.14^{+0.05}_{-0.07}$& --\\ 
&$0.36\pm0.14$&$0.02\pm0.01$&$0.00\pm0.04$
&$1.01^{+0.17}_{-0.14}$&$0.08^{+0.05}_{-0.03}$&$-0.18^{+0.06}_{-0.07}$&  + \\ 
\hline
\multirow{2}{*}{\makecell[c]{0.75}} &\detwo&$0.04^{+0.05}_{-0.02}$&$0.00\pm0.07$&
\bfafwo & $0.10^{+0.11}_{-0.05}$&$-0.20^{+0.09}_{-0.12}$& --\\ 
&\detwith&$0.04^{+0.04}_{-0.02}$&$0.00\pm0.07$&
$\mathbf{1.17^{+0.30}_{-0.21}}$&$\mathbf{0.13^{+0.12}_{-0.06}}$&$\mathbf{-0.26^{+0.09}_{-0.13}}$&  + \\ 
\hline
\multirow{2}{*}{\makecell[c]{0.96}} & \multirow{2}{*}{\makecell[c]{$0.27^{+0.23}_{-0.22}$}}
& \multirow{2}{*}{\makecell[c]{$0.02^{+0.03}_{-0.01}$}}& \multirow{2}{*}{\makecell[c]{$-0.08\pm0.06$}}&
$0.73^{+0.34}_{-0.22}$ & $0.05^{+0.11}_{-0.03}$&$-0.25^{+0.08}_{-0.14}$& --\\ 
& & & &$1.23^{+0.52}_{-0.29}$&$0.16^{+0.31}_{-0.10}$&$-0.35^{+0.13}_{-0.23}$&  + \\ 
\hline
\hline
\end{tabular}
}
\end{center}
\end{minipage}
\label{tab:fit_res}
\end{table*}

One can also consider the constraints on the \lr---\lx\ correlation that can be derived using the 2016 outburst measurements alone.  Unfortunately, given the relatively small number of hard-state observations (4 in total with HR $>0.75$), the {\tt LINMIX\_ERR} method does not converge.  Alternatively, we fit the 2016 outburst data points using a simple least-squares method, in which the upper limits were treated as 3-$\sigma$ detections.  This resulted in a best-fit slope $\beta = 0.82$, which is roughly speaking a 3-$\sigma$ lower-limit on the steepness of the slope.  This indicates that the 2016 data are consistent with the slope inferred from modelling all outbursts together.

\section{Discussion}
\label{sec:disc}

Using the VLA, we have performed high-sensitivity, quasi-simultaneous radio---X-ray observations of \aql\ that probe an order of magnitude deeper in radio luminosity compared to previous studies.  This allows us to meaningfully constrain the \lr---\lx\ correlation of \aql\ over 2 dex, down to low X-ray luminosities ($6 \times 10^{34}$\,erg\,s$^{-1} <$ \lx\ $< 3 \times 10^{35}$\,erg\,s$^{-1}$), as the source was fading from outburst.

ATCA detections of radio emission from \aql\ during the same outburst indicate that the jet was present throughout the outburst when \lx\ $> 5 \times 10^{36}$\,erg\,s$^{-1}$ \citep{diaztrigo2018}. The last ATCA detection (with HR $= 0.64$) was performed just 6 days before our first VLA observation.  Thus our VLA non-detections are indeed indicative of a fading jet.

Here we discuss the derived constraints on the \lr---\lx\ correlation for \aql, compare its behaviour to other NS-LMXBs, and discuss the motivation for future radio---X-ray observations.

\subsection{\lr---\lx\ correlation in {\aql}}

\aql's jet behaviour during the 2016 outburst was possibly somewhat unusual: unlike what was previously observed \citep{MJ2010}, the radio emission was not seen to quench at X-ray luminosities above $10^{37}$\,erg\,s$^{-1}$ (in the soft X-ray state), and rather reached a record brightness of 810\,$\mu$Jy at 5.5\,GHz.  However, the sparse observational sampling may mean that the quenching period was simply missed.  Noteworthy is that the 2016 outburst is the only high-luminosity outburst \citep[reaching about factor of 5 higher X-ray luminosity; see][]{gungor2014} of \aql\ for which both radio and X-ray observations were performed, and thus they probed unprecedentedly high X-ray luminosities.  Nonetheless, and as discussed before, the radio luminosities of the two observations performed during the (relatively) hard state (at \lx $\sim 10^{36-37}$ erg\,s$^{-1}$) are consistent with previous observations from past outbursts.  This suggests that it is appropriate to combine the 2016 outburst measurements with those of previous outbursts to study the \lr---\lx\ correlation of \aql\ using the largest available data set.

\subsubsection{Modelling \lr---\lx\ with a single power-law}

We collected all archival radio and X-ray measurements prior to 2016 and fit these with a slope $\beta = $ \bfafwo.  As we discuss in \S\ref{sec:results} and summarise in Table~\ref{tab:fit_res}, the slope is highly dependent on the inclusion of radio upper limits.  While we exclude soft-state measurements, the slope depends only slightly on the chosen hardness ratio threshold.  Ultimately, we find a slope that is consistent within uncertainties with the $\beta = 0.76^{+0.14}_{-0.15}$ of \citet{Tetarenko2016}, despite the fact they used data from fewer outbursts and only included detections in their fitting.  Based on this slope, \citet{Tetarenko2016} suggested that \aql\ is consistent with being radiatively inefficient, and similar to BH-LMXBs.  However, it is hard to draw robust conclusions from the previously available data because of the limited number of measurements and the fact that they span only $\sim 1$\,dex in radio and X-ray luminosity.

The 2016 measurements probe $\sim 2$\,dex in luminosity and provide significant new constraints.  Including the 2016 outburst and all archival data, the resulting slope is $\beta =$ \bfafwith\ for HR $> 0.75$. This is steeper than what has been reported in the previous studies mentioned above, but also consistent with what we obtained using only archival data.  Given the small number of available observations, it is not possible to perform a robust separate fit to the 2016 outburst data alone, but these data are consistent with the global fit. In the framework of models in which jet power is a fixed fraction of the accretion power, this newly inferred slope $\beta =$ \bfafwith\ indicates that \aql\ has a more radiatively efficient accretion inflow than what was previously reported.

Recently, \citet{Qiao2019} also used pre-2016 archival data of \aql\ and found a shallow ($\beta \sim 0.4$) correlation between the radio and X-ray luminosities. These results are comparable to ours if we omit radio upper limits in our fitting (see Table~\ref{tab:fit_res}). However, as we demonstrate, the inclusion of upper limits results in a steeper $\beta=$\bfafwo. In their work, \citet{Qiao2019} attempt to explain their fitted correlation of $\beta\sim0.4$ with a coupled advection-dominated accretion (ADAF)-jet model \citep{Yuan2005,Done2007}, where they assume that the accretion flow is always radiatively efficient for NSs, but that the fraction of the jet power to the accretion power is not constant. They assume that $\beta\sim0.4$ correlation holds down to low X-ray luminosities (\lx\ $\sim 10^{34-36}$\,erg\,s$^{-1}$) and predict the corresponding radio luminosities to be \lr\ $\sim 1-4 \times 10^{28}$\,erg\,s$^{-1}$. Our VLA upper limits on radio luminosity are about an order of magnitude fainter than was predicted by \citet{Qiao2019}. Therefore, such a scenario is inconsistent with our new 2016 observations.  Moreover, the $\beta =$\bfafwo\ that we find for archival data (including upper limits) suggests that the scenario is also inconsistent with the observations from previous outbursts.

\begin{figure*}
\centering 
\includegraphics[width=\textwidth]{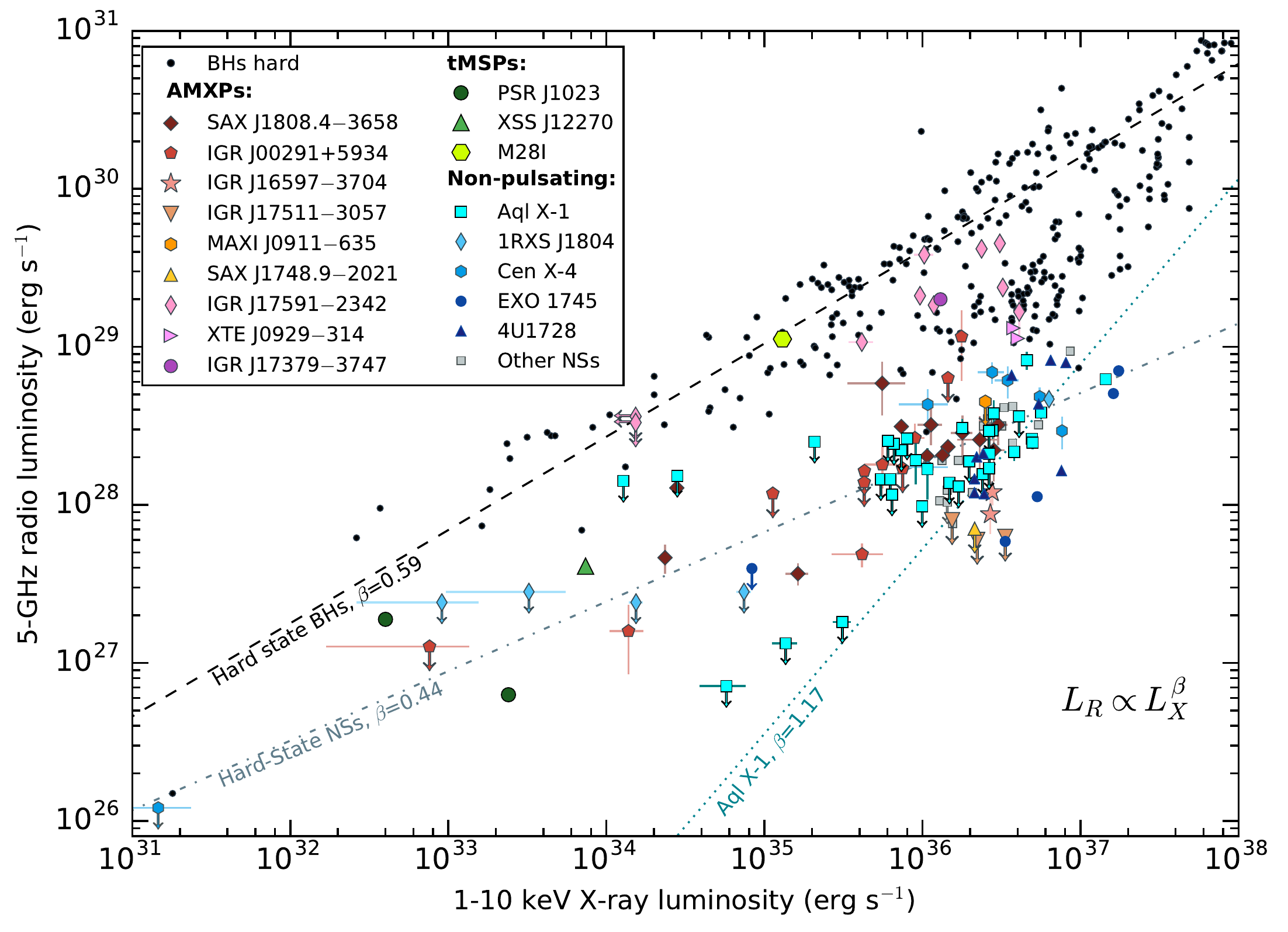}

\caption{X-ray ($1-10$\,keV) luminosity versus radio (5\,GHz)
  luminosity for hard state (HR $\gtrsim$ 0.75) BH- and NS-LMXBs. Black circles represent BH-LMXBs; Blue symbols represent non-pulsating NS-LMXBs for which many observations have been obtained and grey squares represent other non-pulsating NSs; red, pink and yellow symbols represent AMXPs and green symbols represent tMSPs. Individual sources within groups are shown in different symbols and colour tones. Data points are taken from: \citet{Bahramian2018} for BHs hard; \citet{TUD2009}, \citet{MJ2010} and \citet{diaztrigo2018} for \aql; \citet{gusinskaia2017} for 1RXS~J180408.9$-$342058; \citet{Tetarenko2016} for EXO~1745$-$248; \citet{MIG2003} for 4U1728$-$34; \citet{Papitto2013} for M28I; \citet{Hill2011} for XSS~J12270$-$4859; \citet{Bogdanov2018} for PSR~J1023+0038; \citet{Tudor2017} for SAX~J1808.4$-$3658, IGR~J00291$-$5934, and IGR~J17511$-$3057; \citet{Tetarenko2018} for IGR~J16597$-$3704, \citet{Tudor2016ATel} for MAXI~J0911$-$635; \citet{MJ2010ATel}, \citet{Tetarenko2017ATel} for SAX~J1748.9$-$2021; \citet{Russell2018} and \citet{Gusinskaia2019_igr} for IGR~J17591$-$2342; \citet{Migliari2011} for XTE~J0929$-$314; \citet{J17379_ATel1148} for IGR~J17379$-$3747; \citet{MIGFEN2006} for other NSs. Correlation tracks for hard state BHs (dashed black line) and NSs (dash-dotted grey line) are defined in \citet{Gallo2018}; light blue dotted line represents the result of our prefered fit for \aql. }\label{fig:lrlx_all}

\end{figure*}

\subsubsection{Modelling \lr---\lx\ with a low-luminosity radio cutoff}

Instead of fitting all data with an assumed single power-law correlation, we can also consider a model in which the radio brightness suddenly drops below a critical X-ray luminosity.  This could be the case if a minimal accretion rate is necessary to sustain a steady jet.

The position of our new VLA observations of \aql\ in the \lr---\lx\ diagram may indicate a sudden step in radio luminosity, rather than a smooth decay. Indeed, during 2016 outburst, the X-ray luminosity at the time of the last ATCA detection and the first VLA upper-limit is different by a factor of 14, while the radio luminosity drops by more than a factor of 24. Furthermore, our first VLA non-detection (with \lx\ $\sim 3 \times 10^{35}$erg\,s$^{-1}$, \lr\ $\leq 1.5\times 10^{27}$erg\,s$^{-1}$) is different from the lowest X-ray luminosity archival data point (\lx $\sim 5.5 \times 10^{35}$erg\,s$^{-1}$, \lr $\leq 1.4\times 10^{28}$erg\,s$^{-1}$) by approximately a factor of 2 in X-ray luminosity, but by more than an order of magnitude in radio luminosity. This again assumes that we can compare separate outbursts.

Given the above, it may be questionable to impose a single power law over the entire \lx\ regime between \lx\ $\sim 6 \times 10^{34}$\,erg\,s$^{-1}$ and \lx\ $\sim 10^{37}$\,erg\,s$^{-1}$. Instead, it could be that \aql\ is switching between two physically different accretion modes around some critical X-ray luminosity, above which \lr---\lx\ follows a shallow $\beta \sim 0.8$ correlation, and below which the jet switches off (or becomes undetectable at $5-10$\,GHz) or does not follow a power-law correlation at all over this broad luminosity range.  Dense sampling of the outburst decay is needed to explore this possibility.

\subsection{Comparison to other NS-LMXBs}

We now compare \aql\ to other NS-LMXBs with quasi-simultaneous radio and X-ray observations (Figure~\ref{fig:lrlx_all}, and associated references).

The best-studied regime is at high X-ray luminosities (\lx\ $> 10^{36}$\,erg\,s$^{-1}$), where over a dozen systems have been observed. These show a broad range of radio luminosities for comparable X-ray luminosity.  \aql\ is one of the fainter systems, being e.g. $\sim 15\times$ less bright compared to IGR~J17591$-$2342 \citep{Russell2018}. The origin of this large spread in radio luminosity is unclear.  Presumably it could depend on viewing geometry, the magnetic field strength of the neutron star, its spin rate, and the inclination of the magnetic axis with respect to the orbital angular momentum.  Determining which of these (or other) physical parameters are most relevant requires additional observations and numerical simulations.  To date, the larger the number of NS-LMXBs that have been studied, the larger the diversity in their \lr---\lx\ behaviour.  \aql\ has shown intermittent pulsations for a period of only 150\,s \citep{CAS2008}.  In the (perhaps unlikely) situation that it shows pulsations for a longer period during a future outburst, simultaneous radio observations can determine whether this has a demonstrable effect on the observed \lr---\lx\ behaviour.  The same argument applies to other intermittent AMXPs.

The situation is even more unclear at intermediate X-ray luminosities ($10^{34}$\,erg\,s$^{-1} <$ \lx\ $<10^{36}$\,erg\,s$^{-1}$), where only 5 systems have been observed, and only two detected.  Both detected systems are AMXPs (SAX~J1808.4$-$3658 and IGR~J00291$-$5934), which led \citet{Tudor2017} to contemplate why these are surprisingly bright and apparently launch jets at relatively low X-ray luminosity.  Our strong radio upper limits in this X-ray luminosity range for \aql\ show that it is certainly dimmer than the two detected AMXPs and consistent with the non-detections of two non-pulsating NS-LMXB systems (1RXS~J180408.9$-$342058 and EXO~1745$-$248).  However, given the large spread in radio luminosity at higher X-ray luminosity, the interpretation is unclear.  It is too early to state that AMXPs are brigher on average compared to non-pulsating NS-LMXBs.  Indeed, \citet{Gallo2018} concluded that there is no strong evidence for a systematic difference between AMXPs and non-pulsating NS-LMXBs.  Here too, many more NS-LMXBs need to be observed, though this is challenging given their low radio brightness and will require using the most sensitive recently commissioned (MeerKAT) and future (Square Kilometre Array) radio telescopes.

Finally, at low X-ray luminosities (\lx\ $<10^{34}$\,erg\,s$^{-1}$), typically referred to as ``quiescent'' state, the only radio-detected NS systems are the tMSPs PSR~J1023+0038 and XSS~J12270$-$4859.  As previously discussed ($\S1$), their radio---X-ray behaviour is demonstrably different compared to NS-LMXBs observed at higher accretion rate, and they are likely in a fundamentally different accretion regime.  It would be very interesting to know whether \aql\ enters a similar regime during quiescence; however, given that it is at least three times further away compared to PSR~J1023+0038, its expected radio flux density may be only $\sim 10$\,$\mu$Jy.  PSR~J1023+0038 showed anti-correlated radio---X-ray brightness variations in which the radio flux density changed by a factor of $\sim 3$ between X-ray high and low modes.  A similar, strictly simultaneous radio---X-ray campaign on \aql\ during quiescence would likely require the sensitivity of future instruments like the Square Kilometre Array and Athena in order to be successful.  However, Cen X-4 is similarly nearby to PSR~J1023+0038 and remains undetected in radio \citep{Tudor2017}.  This suggests that not all NS-LMXBs enter a low-luminosity accretion regime similar to that observed in the tMSPs.  Possible explanations include the neutron star magnetic field strength or spin rate.

Despite the complications described above, there have been attempts to describe the population of NS-LMXBs with a single \lr---\lx\ relation. \citet{Gallo2018} studied the \lr---\lx\ correlation of all NS-LMXBs combined as well as for different NS-LMXB classes. They found a correlation for the combined sample of NS-LMXBs with a slope $\beta = 0.44^{+0.05}_{-0.04}$ and $\beta = 0.71^{+0.11}_{-0.09}$ for only atoll-type NS-LMXBs, both suggestive of radiatively inefficient accretion. Surprisingly, this result does not match what was so far observed in individual atoll-type NS-LMXBS for which the \lr---\lx\ correlation was studied in detail. As mentioned earlier, power-law slopes of $\beta = 1.5\pm0.2$, $\beta = 1.68^{+0.10}_{-0.09}$ and $\beta=$\bfafwith\ were found for 4U~1728$-$34, EXO~1745$-$248 and \aql, respectively --- all consistent with radiatively efficient accretion. This argues that differences in the behaviour of individual sources --- such as inconsistent \lr---\lx\ relationships \citep{Tudor2017} or radio luminosity offset \citep{Tetarenko2016} --- are significant.

\subsection{Conclusions}

In summary, we have acquired quasi-simultaneous radio---X-ray observations that allow us to provide the strongest-ever constraints on the radio brightness of \aql\ at low X-ray luminosities (\lx\ $<10^{36}$\,erg\,s$^{-1}$).  We combine these new measurements with an exhaustive re-analysis of all available radio---X-ray data from 7 previous outbursts of \aql.  When modelling this combined data set with a single power-law, we find a slope of \bfafwith, which is steeper than previously reported results in the literature and is consistent with radiatively efficient accretion.  The inclusion of radio upper limits, which were ignored in the fits of previous studies, already lead to a steep inferred slope (\bfafwo) even if the 2016 outburst is excluded.

\section*{Acknowledgements}

N.V.G. acknowledges funding from NOVA.  J.W.T.H. acknowledges funding
from an NWO Vidi fellowship and from the European Research Council
under the European Union's Seventh Framework Programme (FP/2007-2013)
/ ERC Starting Grant agreement nr. 337062 (``DRAGNET''). The Australia Telescope Compact Array is part of the Australia Telescope National Facility, which is funded by the Australian Government for operation as a National Facility managed by CSIRO. C.A.M.-J. is the recipient of an Australian Research Council Future Fellowship (FT140101082), funded by the Australian government. N.D. is supported by a Vidi grant from NWO.

\bibliographystyle{mnras}
\bibliography{allbib}

\onecolumn
\newpage
\appendix

\section{}



\LTcapwidth=\textwidth

\begin{center}
\renewcommand{\arraystretch}{1.2}
\begin{longtable}{@{\extracolsep{-0.3pt}}l|cccccc|ccc@{}}
\caption[All observations]{All radio and X-ray quasi-simultaneous observations used in this study}\label{tab:appendix} \\
\hline\hline
\multirow{3}{*}{\makecell[c]{Out-\\burst}}  & \multicolumn{3}{c}{Radio (VLA$^{a}$)} & \multicolumn{3}{c|}{X-ray ({\it RXTE}-PCA)} & Radio &  \multicolumn{2}{c}{X-ray}\\
  \cline{2-4}
  \cline{5-8}
  \cline{9-10}
 & \makecell[c]{ MJD } & \makecell[c]{$S_{\nu}$ \\ ($\upmu$Jy)} & \makecell[c]{Tel. $\nu$\\ \tiny{(GHz)}}  & \makecell[c]{MJD } & \makecell[c]{Unabsorbed \\ flux  $\times 10^{-10}$\\($\mathrm{erg\,s^{-1}\,cm^{-2}}$)}   &  \makecell[c]{HR } &  \makecell[c]{Lum.$^b$ \\ $\times 10^{28}$\\($\mathrm{erg\,s^{-1}}$)} & \makecell[c]{Lum.$^b$ \\ $\times 10^{36}$\\($\mathrm{erg\,s^{-1}}$)} &  \makecell[c]{HR } \\
\hline

\multirow{3}{*}{\makecell[c]{Mar.\\ 2002}}
& 52334.620 & < 69 & 8.46 & \makecell[c]{52332.922\\ 52335.036} & \makecell[c]{33.74$\pm$0.19\\ 75.41$\pm$0.23} & \makecell[c]{0.47$\pm$0.01\\ 0.45$\pm$0.01} &  < 0.84 & 15.59$\pm$1.38 & 0.46$\pm$0.03  \\ 
& 52355.670 & 179$\pm$26 & 8.46 & \makecell[c]{52354.851\\ 52355.838} & \makecell[c]{19.33$\pm$0.07\\ 10.06$\pm$0.04} & \makecell[c]{0.43$\pm$0.01\\ 0.70$\pm$0.01} &  2.17$\pm$0.31 & 2.72$\pm$0.10 & 0.64$\pm$0.02  \\ 
\hline
\multirow{10}{*}{\makecell[c]{Apr.\\ 2004}}
& 53085.560 & < 174 & 8.46 & 53085.933& 10.93$\pm$0.04& 1.11$\pm$0.01&  < 2.11 & 2.65$\pm$0.01 & 1.11$\pm$0.01  \\ 
& 53144.500 & 179$\pm$22 & 8.46 & \makecell[c]{53143.651\\ 53144.627} & \makecell[c]{15.56$\pm$0.07\\ 15.73$\pm$0.06} & \makecell[c]{1.12$\pm$0.01\\ 1.14$\pm$0.01} &  2.17$\pm$0.27 & 3.81$\pm$0.12 & 1.14$\pm$0.02  \\ 
& 53151.420 & 216$\pm$19 & 8.46 & \makecell[c]{53150.922\\ 53151.707} & \makecell[c]{20.66$\pm$0.10\\ 20.18$\pm$0.09} & \makecell[c]{1.06$\pm$0.01\\ 1.07$\pm$0.01} &  2.62$\pm$0.23 & 4.93$\pm$0.31 & 1.07$\pm$0.02  \\ 
& 53162.370 & 205$\pm$19 & 8.46 & \makecell[c]{53160.757\\ 53162.668} & \makecell[c]{15.96$\pm$0.06\\ 21.50$\pm$0.08} & \makecell[c]{1.13$\pm$0.01\\ 1.11$\pm$0.01} &  2.48$\pm$0.23 & 4.97$\pm$0.32 & 1.11$\pm$0.02  \\ 
& 53170.350 & 274$\pm$22 & 8.46 & 53170.404& 55.89$\pm$0.17& 0.47$\pm$0.01&  3.32$\pm$0.27 & 13.54$\pm$0.04 & 0.47$\pm$0.01  \\ 
& 53176.290 & < 114 & 8.46 & \makecell[c]{53174.408\\ 53176.919} & \makecell[c]{17.23$\pm$0.06\\ 4.32$\pm$0.02} & \makecell[c]{0.56$\pm$0.01\\ 1.08$\pm$0.01} &  < 1.38 & 1.48$\pm$0.19 & 0.91$\pm$0.04  \\ 
& 53183.290 & < 126 & 8.46 & \makecell[c]{53182.753\\ 53184.720} & \makecell[c]{0.16$\pm$0.01\\ 0.05$\pm$0.00} & \makecell[c]{1.20$\pm$0.18\\ 1.90$\pm$0.30} &  < 1.53 & 0.03$\pm$0.00 & 1.36$\pm$0.34  \\ 
& 53187.230 & < 117 & 8.46 & 53186.758& 0.05$\pm$0.00& 1.45$\pm$0.34&  < 1.42 & 0.01$\pm$0.00 & 1.45$\pm$0.34  \\ 
\hline
\multirow{4}{*}{\makecell[c]{May\\ 2005}}
& 53465.540 & 245$\pm$29 & 4.86 & \makecell[c]{53465.073\\ 53465.660} & \makecell[c]{11.60$\pm$0.06\\ 11.40$\pm$0.04} & \makecell[c]{1.08$\pm$0.01\\ 1.13$\pm$0.01} &  2.97$\pm$0.35 & 2.77$\pm$0.08 & 1.12$\pm$0.01  \\ 
& 53472.500 & 317$\pm$30 & 4.86 & \makecell[c]{53472.221\\ 53472.810} & \makecell[c]{23.25$\pm$0.10\\ 23.25$\pm$0.08} & \makecell[c]{1.06$\pm$0.01\\ 1.14$\pm$0.01} &  3.84$\pm$0.36 & 5.63$\pm$0.34 & 1.10$\pm$0.02  \\ 
& 53494.520 & < 156 & 4.86 & 53493.789& 8.15$\pm$0.04& 1.10$\pm$0.01&  < 1.89 & 1.97$\pm$0.01 & 1.10$\pm$0.01  \\ 
& 53506.480 & < 207 & 4.86 & 53505.848& 0.86$\pm$0.01& 1.12$\pm$0.03&  < 2.51 & 0.21$\pm$0.01 & 1.12$\pm$0.03  \\ 
\hline
\multirow{9}{*}{\makecell[c]{Aug.\\ 2006}}
& 53949.404 & 252$\pm$64 & 4.86 & 53949.415& 6.92$\pm$0.03& 0.70$\pm$0.01&  3.05$\pm$0.78 & 1.68$\pm$0.01 & 0.70$\pm$0.01  \\ 
& 53950.209 & 300$\pm$35 & 8.46 & \makecell[c]{53949.415\\ 53950.347} & \makecell[c]{6.92$\pm$0.03\\ 10.19$\pm$0.04} & \makecell[c]{0.70$\pm$0.01\\ 0.60$\pm$0.01} &  3.63$\pm$0.42 & 2.33$\pm$0.08 & 0.61$\pm$0.02  \\ 
& 53951.062 & 180$\pm$42 & 8.46 & \makecell[c]{53950.347\\ 53951.445} & \makecell[c]{10.19$\pm$0.04\\ 11.09$\pm$0.04} & \makecell[c]{0.60$\pm$0.01\\ 0.67$\pm$0.01} &  2.18$\pm$0.51 & 2.61$\pm$0.21 & 0.64$\pm$0.03  \\ 
& 53954.377 & 313$\pm$67 & 4.86 & \makecell[c]{53954.204\\ 53955.973} & \makecell[c]{12.19$\pm$0.05\\ 7.59$\pm$0.04} & \makecell[c]{0.86$\pm$0.01\\ 1.00$\pm$0.01} &  3.79$\pm$0.81 & 2.82$\pm$0.11 & 0.87$\pm$0.02  \\ 
& 53956.269 & < 108 & 8.46 & \makecell[c]{53955.973\\ 53958.059} & \makecell[c]{7.59$\pm$0.04\\ 4.25$\pm$0.02} & \makecell[c]{1.00$\pm$0.01\\ 1.05$\pm$0.01} &  < 1.31 & 1.69$\pm$0.11 & 1.01$\pm$0.02  \\ 
& 53958.354 & < 81 & 8.46 & \makecell[c]{53958.059\\ 53959.634} & \makecell[c]{4.25$\pm$0.02\\ 3.56$\pm$0.02} & \makecell[c]{1.05$\pm$0.01\\ 1.03$\pm$0.01} &  < 0.98 & 1.00$\pm$0.06 & 1.05$\pm$0.02  \\ 
& 53965.282 & < 120 & 8.46 & \makecell[c]{53964.610\\ 53965.460} & \makecell[c]{3.11$\pm$0.02\\ 2.07$\pm$0.01} & \makecell[c]{1.07$\pm$0.02\\ 1.07$\pm$0.02} &  < 1.45 & 0.55$\pm$0.02 & 1.07$\pm$0.03  \\  
\hline
{\makecell[c]{May\\ 2007}}
& 54242.589 & < 243 & 8.46 & \makecell[c]{54242.409\\ 54243.086} & \makecell[c]{10.39$\pm$0.04\\ 12.38$\pm$0.05} & \makecell[c]{1.10$\pm$0.01\\ 1.06$\pm$0.01} &  < 2.94 & 2.64$\pm$0.11 & 1.09$\pm$0.02  \\ 
\hline
\multirow{4}{*}{\makecell[c]{Oct.\\ 2007}}
& 54376.003 & 158$\pm$47 & 8.46 & 54376.994& 3.74$\pm$0.02& 1.06$\pm$0.02&  1.91$\pm$0.57 & 0.91$\pm$0.00 & 1.06$\pm$0.02  \\ 
& 54379.181 & < 96 & 8.46 & \makecell[c]{54377.325\\ 54379.867} & \makecell[c]{3.35$\pm$0.02\\ 2.43$\pm$0.02} & \makecell[c]{1.09$\pm$0.01\\ 1.08$\pm$0.02} &  < 1.16 & 0.64$\pm$0.09 & 1.08$\pm$0.04  \\ 
& 54380.865 & < 183 & 8.46 & \makecell[c]{54379.867\\ 54382.364} & \makecell[c]{2.43$\pm$0.02\\ 4.25$\pm$0.02} & \makecell[c]{1.08$\pm$0.02\\ 1.07$\pm$0.01} &  < 2.22 & 0.74$\pm$0.15 & 1.08$\pm$0.05  \\ 
\hline
\multirow{6}{*}{\makecell[c]{Nov.\\ 2009}}
& 55140.125 & < 201 & 8.46 & 55140.666& 2.73$\pm$0.01& 1.12$\pm$0.02&  < 2.43 & 0.66$\pm$0.00 & 1.12$\pm$0.02  \\ 
& 55141.185 & < 210 & 8.46 & \makecell[c]{55140.666\\ 55141.379} & \makecell[c]{2.73$\pm$0.01\\ 2.42$\pm$0.01} & \makecell[c]{1.12$\pm$0.02\\ 1.11$\pm$0.02} &  < 2.54 & 0.60$\pm$0.03 & 1.11$\pm$0.02  \\ 
& 55142.992 & < 120 & 8.46 & \makecell[c]{55142.093\\ 55143.074} & \makecell[c]{1.95$\pm$0.01\\ 2.65$\pm$0.01} & \makecell[c]{1.12$\pm$0.02\\ 1.13$\pm$0.02} &  < 1.45 & 0.63$\pm$0.01 & 1.13$\pm$0.03  \\ 
& 55144.966 & < 129 & 8.46 & \makecell[c]{55144.122\\ 55144.967} & \makecell[c]{7.13$\pm$0.03\\ 9.91$\pm$0.04} & \makecell[c]{1.11$\pm$0.01\\ 1.11$\pm$0.01} &  < 1.56 & 2.40$\pm$0.01 & 1.11$\pm$0.02  \\ 
& 55146.089 & < 141 & 8.46 & \makecell[c]{55146.083\\ 55147.142} & \makecell[c]{10.89$\pm$0.05\\ 11.93$\pm$0.06} & \makecell[c]{1.12$\pm$0.01\\ 1.11$\pm$0.01} &  < 1.71 & 2.64$\pm$0.02 & 1.12$\pm$0.02  \\ 
& 55149.080 & < 300 & 8.46 & \makecell[c]{55148.246\\ 55149.094} & \makecell[c]{13.94$\pm$0.06\\ 16.92$\pm$0.09} & \makecell[c]{1.11$\pm$0.01\\ 1.05$\pm$0.01} &  < 3.63 & 4.09$\pm$0.03 & 1.05$\pm$0.01  \\ 
& 55152.072 & 679$\pm$91 & 8.46 & \makecell[c]{55151.189\\ 55152.170} & \makecell[c]{20.16$\pm$0.08\\ 18.76$\pm$0.07} & \makecell[c]{0.96$\pm$0.01\\ 0.85$\pm$0.01} &  8.23$\pm$1.10 & 4.58$\pm$0.11 & 0.86$\pm$0.01  \\ 
& 55154.000 & 406$\pm$62 & 4.86 & \makecell[c]{55153.080\\ 55154.602} & \makecell[c]{32.10$\pm$0.11\\ 62.98$\pm$0.20} & \makecell[c]{0.69$\pm$0.01\\ 0.46$\pm$0.01} &  4.92$\pm$0.75 & 11.69$\pm$1.47 & 0.54$\pm$0.03  \\ 
\multirow{10}{*}{\makecell[c]{Nov.\\ 2009}}
& 55154.698 & 400$\pm$50 & \makecell[c]{\tiny{EVN} \\5.00} & \makecell[c]{55154.602\\ 55155.043} & \makecell[c]{62.98$\pm$0.20\\ 65.81$\pm$0.21} & \makecell[c]{0.46$\pm$0.01\\ 0.47$\pm$0.01} &  4.85$\pm$0.61 & 15.40$\pm$0.37 & 0.46$\pm$0.01  \\ 
& 55155.810 & 203$\pm$62 & 8.46 & \makecell[c]{55155.043\\ 55156.095} & \makecell[c]{65.81$\pm$0.21\\ 94.28$\pm$0.28} & \makecell[c]{0.47$\pm$0.01\\ 0.45$\pm$0.00} &  2.46$\pm$0.75 & 20.72$\pm$1.29 & 0.45$\pm$0.02  \\ 
& 55156.815 & < 81 & 8.46 & \makecell[c]{55156.095\\ 55157.141} & \makecell[c]{94.28$\pm$0.28\\ 59.35$\pm$0.30} & \makecell[c]{0.45$\pm$0.00\\ 0.47$\pm$0.01} &  < 0.98 & 16.61$\pm$1.17 & 0.47$\pm$0.02  \\ 
& 55157.124 & < 317 & \makecell[c]{\tiny{VLBA} \\8.41} & \makecell[c]{55156.095\\ 55157.141} & \makecell[c]{94.28$\pm$0.28\\ 59.35$\pm$0.30} & \makecell[c]{0.45$\pm$0.00\\ 0.47$\pm$0.01} &  < 3.84 & 14.49$\pm$0.12 & 0.47$\pm$0.01  \\ 
& 55157.865 & 229$\pm$56 & \makecell[c]{\tiny{VLBA} \\8.41} & \makecell[c]{55157.849\\ 55158.973} & \makecell[c]{83.05$\pm$0.26\\ 114.71$\pm$0.34} & \makecell[c]{0.46$\pm$0.01\\ 0.45$\pm$0.00} &  2.77$\pm$0.68 & 20.21$\pm$0.16 & 0.46$\pm$0.01  \\ 
& 55159.104 & < 126 & 8.46 & \makecell[c]{55158.973\\ 55160.150} & \makecell[c]{114.71$\pm$0.34\\ 103.68$\pm$0.31} & \makecell[c]{0.45$\pm$0.00\\ 0.45$\pm$0.00} &  < 1.53 & 27.48$\pm$0.86 & 0.45$\pm$0.02  \\ 
& 55161.061 & < 189 & \makecell[c]{\tiny{VLBA} \\8.41} & \makecell[c]{55160.150\\ 55161.066} & \makecell[c]{103.68$\pm$0.31\\ 116.42$\pm$0.35} & \makecell[c]{0.45$\pm$0.00\\ 0.46$\pm$0.00} &  < 2.29 & 28.19$\pm$0.17 & 0.46$\pm$0.01  \\ 
& 55162.939 & < 140 & \makecell[c]{\tiny{VLBA} \\8.41} & \makecell[c]{55162.047\\ 55163.028} & \makecell[c]{118.39$\pm$0.35\\ 110.79$\pm$0.33} & \makecell[c]{0.47$\pm$0.01\\ 0.46$\pm$0.00} &  < 1.70 & 27.00$\pm$0.61 & 0.46$\pm$0.01  \\ 
& 55163.945 & < 96 & 8.46 & \makecell[c]{55163.943\\ 55165.120} & \makecell[c]{86.40$\pm$0.28\\ 63.60$\pm$0.24} & \makecell[c]{0.44$\pm$0.01\\ 0.47$\pm$0.01} &  < 1.16 & 20.92$\pm$0.11 & 0.44$\pm$0.01  \\ 
& 55172.005 & 253$\pm$36 & 8.46 & \makecell[c]{55171.072\\ 55172.052} & \makecell[c]{9.71$\pm$0.05\\ 7.27$\pm$0.04} & \makecell[c]{0.90$\pm$0.01\\ 0.95$\pm$0.01} &  3.06$\pm$0.44 & 1.79$\pm$0.03 & 0.95$\pm$0.02  \\ 
& 55173.088 & 139$\pm$50 & 8.46 & \makecell[c]{55173.033\\ 55174.278} & \makecell[c]{4.52$\pm$0.02\\ 2.90$\pm$0.02} & \makecell[c]{1.05$\pm$0.01\\ 1.09$\pm$0.02} &  1.68$\pm$0.61 & 1.07$\pm$0.02 & 1.05$\pm$0.02  \\ 
& 55173.918 & < 217 & \makecell[c]{\tiny{VLBA} \\8.41} & \makecell[c]{55173.033\\ 55174.278} & \makecell[c]{4.52$\pm$0.02\\ 2.90$\pm$0.02} & \makecell[c]{1.05$\pm$0.01\\ 1.09$\pm$0.02} &  < 2.63 & 0.80$\pm$0.06 & 1.08$\pm$0.02  \\ 
\hline
\multirow{10}{*}{\makecell[c]{Sep.\\ 2016}}
& 57602.543 & 515$\pm$26 & \makecell[c]{\tiny{ATCA} \\5.50} & 57603.137& 59.53$\pm$1.04& 0.86$\pm$0.09&  6.24$\pm$0.31 & 14.42$\pm$0.25 & 0.86$\pm$0.09  \\ 
& 57604.509 & 498$\pm$19 & \makecell[c]{\tiny{ATCA} \\5.50} & \makecell[c]{57603.137\\ 57605.111} & \makecell[c]{59.53$\pm$1.04\\ 246.33$\pm$2.12} & \makecell[c]{0.86$\pm$0.09\\ 0.29$\pm$0.03} &  6.03$\pm$0.23 & 38.70$\pm$4.85 & 0.40$\pm$0.09  \\ 
& 57607.599 & 810$\pm$19 & \makecell[c]{\tiny{ATCA} \\5.50} & \makecell[c]{57606.140\\ 57607.726} & \makecell[c]{356.41$\pm$3.25\\ 313.12$\pm$5.30} & \makecell[c]{0.44$\pm$0.04\\ 0.44$\pm$0.04} &  9.81$\pm$0.23 & 76.65$\pm$2.33 & 0.44$\pm$0.06  \\ 
& 57650.354 & 259$\pm$13 & \makecell[c]{\tiny{ATCA} \\5.50} & 57650.597& 18.73$\pm$0.30& 0.64$\pm$0.06&  3.14$\pm$0.16 & 4.54$\pm$0.07 & 0.64$\pm$0.06  \\ 
& 57657.031 & < 15 & 10.00 & \makecell[c]{57656.293\\ 57657.095} & \makecell[c]{1.95$\pm$0.14\\ 1.24$\pm$0.09} & \makecell[c]{1.16$\pm$0.12\\ 1.16$\pm$0.12} &  < 0.18 & 0.31$\pm$0.04 & $>$1  \\ 
& 57658.031 & < 11 & 10.00 & \makecell[c]{57657.095\\ 57658.558} & \makecell[c]{1.24$\pm$0.09\\ 0.36$\pm$0.04} & \makecell[c]{1.16$\pm$0.12\\ 0.95$\pm$0.10} &  < 0.13 & 0.14$\pm$0.02 & $>$1  \\ 
& 57659.181 & < 6 & 10.00 & \makecell[c]{57658.558\\ 57659.767} & \makecell[c]{0.36$\pm$0.04\\ 0.16$\pm$0.07} & \makecell[c]{0.95$\pm$0.10\\ 1.52$\pm$0.15} &  < 0.07 & 0.06$\pm$0.02 & $>$1 \\ 
\hline

\end{longtable}
\begin{flushleft}{
$^{a}$ Unless specified\\
$^b$ Luminosities are calculated assuming the distance of 4.5 kpc.\\
X-ray energy range is 1--10\,keV.
  }\end{flushleft}
\end{center}

\begin{figure*}
\centering 
\includegraphics[width=17cm]{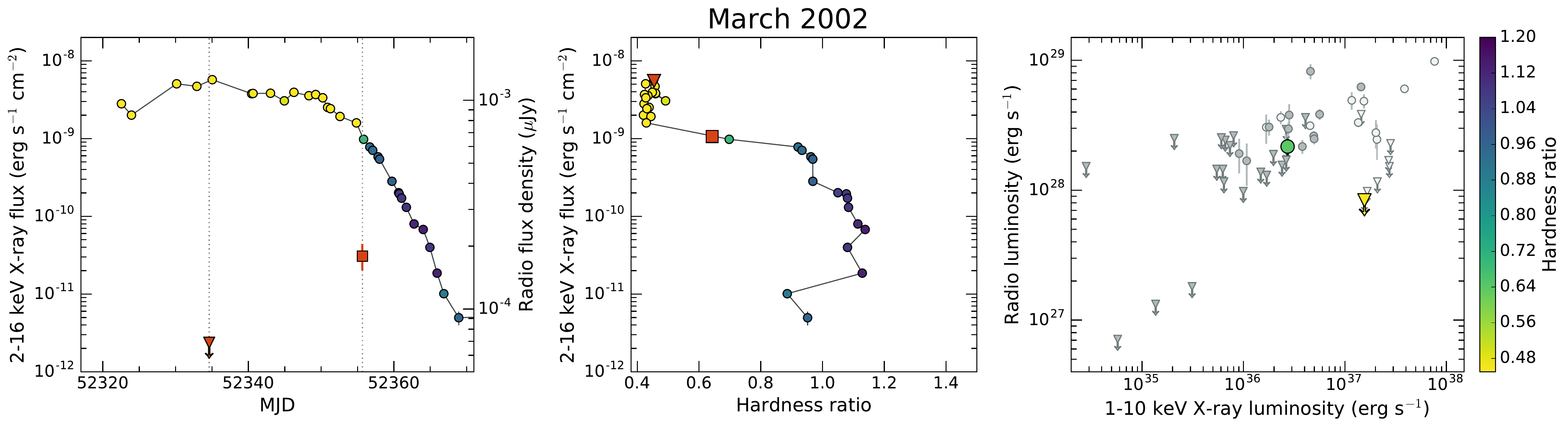}
\caption{Radio and X-ray observations of March 2002 outburst. In each plot, colour scatter of points represents the HR (see colourbar on the right-hand side of this figure) defined as $9.7-16$\,keV flux / $6.0-9.7$\,keV flux. \textbf{\textit{Left}}: Circular symbols represent the {\it RXTE}-PCA ($2-16$ keV)  X-ray light-curve (using the left-hand axis). Red symbols represent the VLA 8.46 GHz radio flux density (using the right-hand axis; squares: detection and triangles: upper-limits). \textbf{\textit{Middle:}} HID of the same outburst. The red symbols indicate at what stage of the outburst the radio observations were taken (squares: detection and triangles: upper-limits). \textbf{\textit{Right:}} Quasi-simultaneous X-ray ($1-10$ keV) luminosity versus radio (5 GHz) luminosity for Aql X-1. Circles and downward-pointing triangles represent radio detections and upper-limits, respectively. Grey and white symbols represent hard (HR $> 0.75$) and soft (HR $< 0.75$) X-ray state observations, respectively. Big coloured symbols indicate observations from March 2002 outburst.}
\vspace{-4mm}
\end{figure*}
\begin{figure*}
\centering 
\includegraphics[width=17cm]{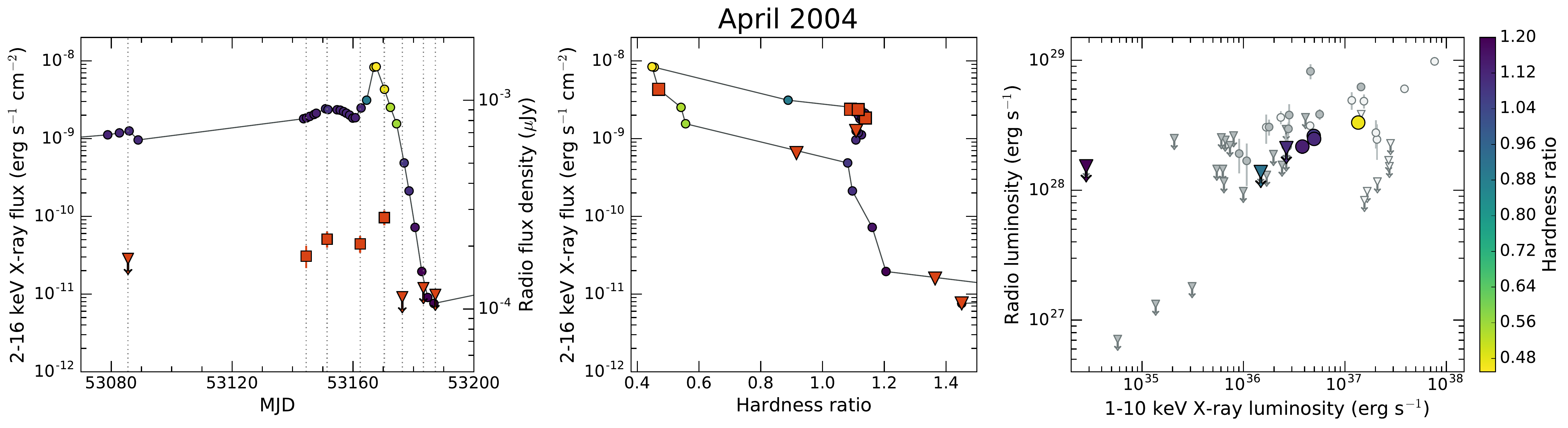}
\caption{Same as Fig. A.1, but for the April 2004 outburst.}
\vspace{-4mm}
\end{figure*}
\begin{figure*}
\centering 
\includegraphics[width=17cm]{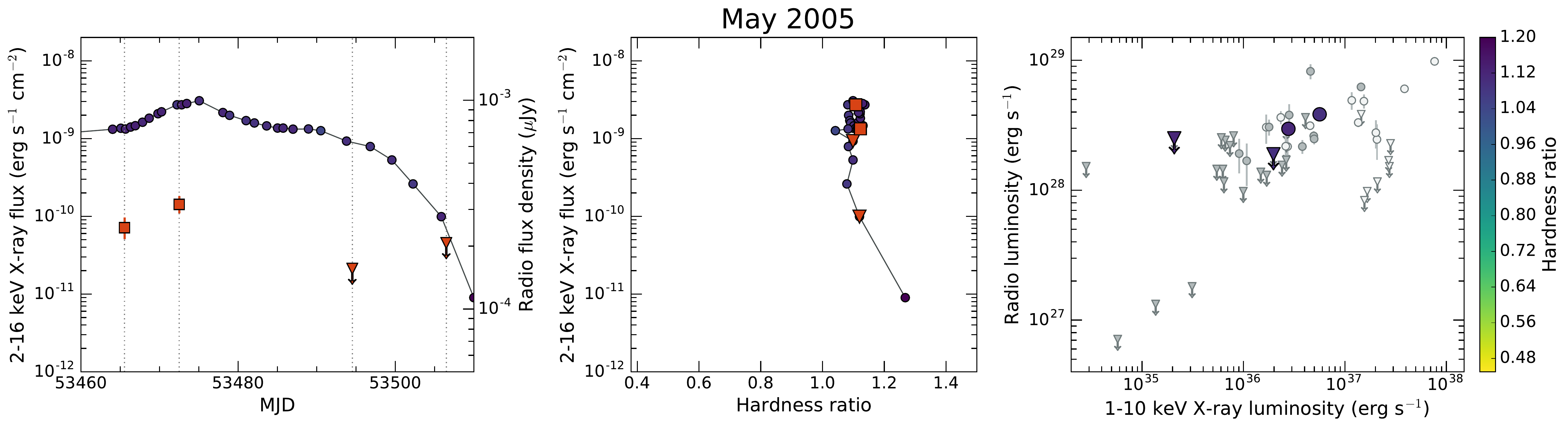}
\caption{Same as Fig. A.1, but for the May 2005 outburst and with radio observations taken at 4.86 GHz.}
\vspace{-4mm}
\end{figure*}
\begin{figure*}
\centering 
\includegraphics[width=17cm]{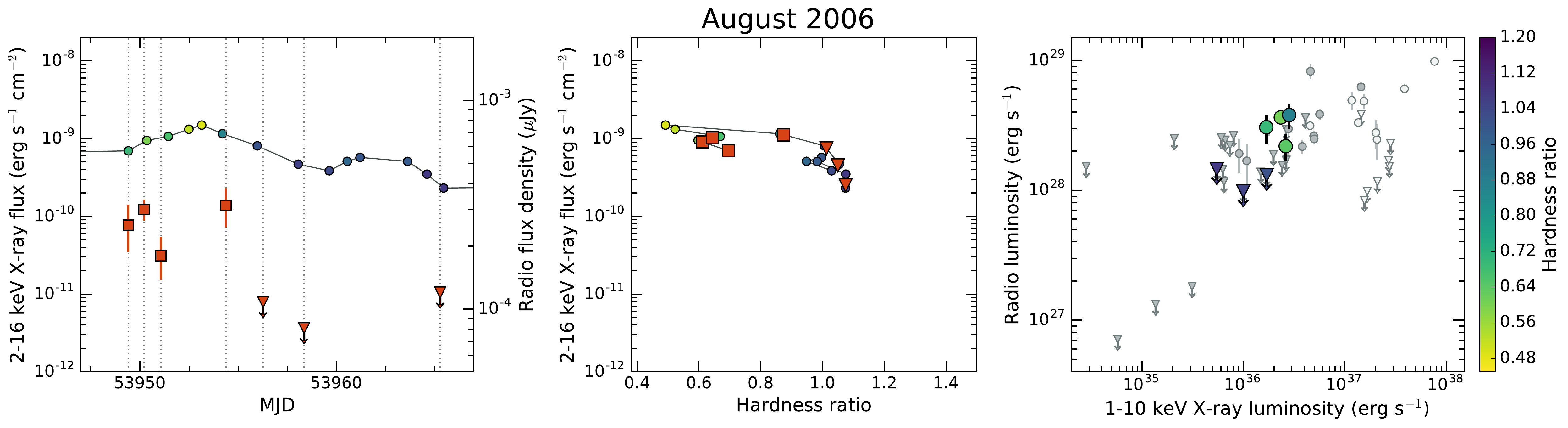}
\caption{Same as Fig. A.1, but for the August 2006 outburst and with radio observations taken at 4.86 GHz and 8.46 GHz.}
\vspace{-4mm}
\end{figure*}
\begin{figure*}
\centering 
\includegraphics[width=17cm]{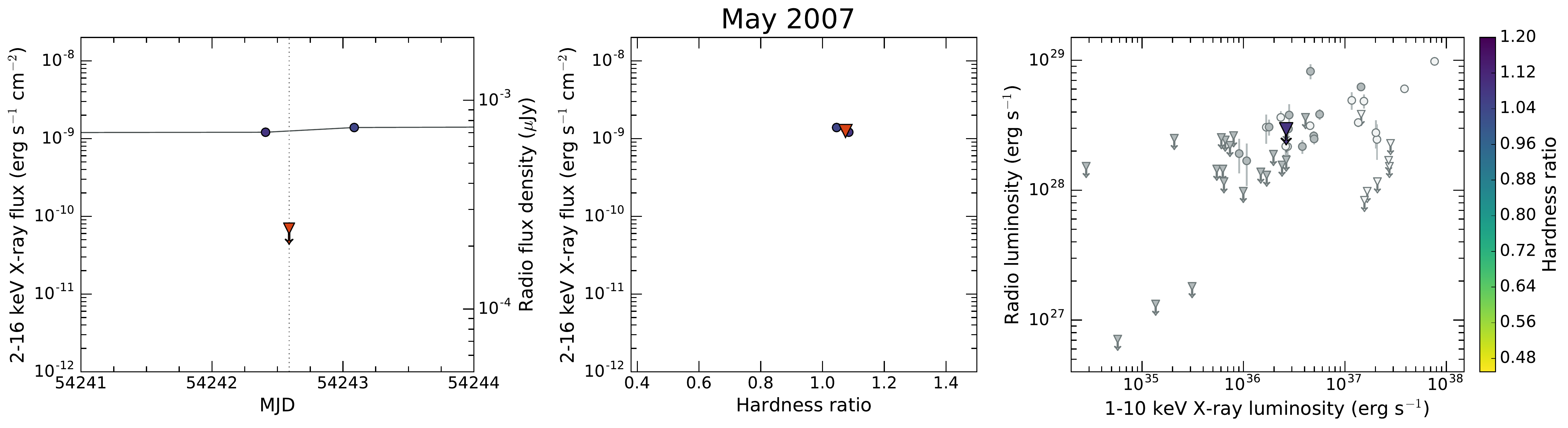}
\caption{Same as Fig. A.1, but for the May 2007 outburst.}
\vspace{-4mm}
\end{figure*}
\begin{figure*}
\centering 
\includegraphics[width=17cm]{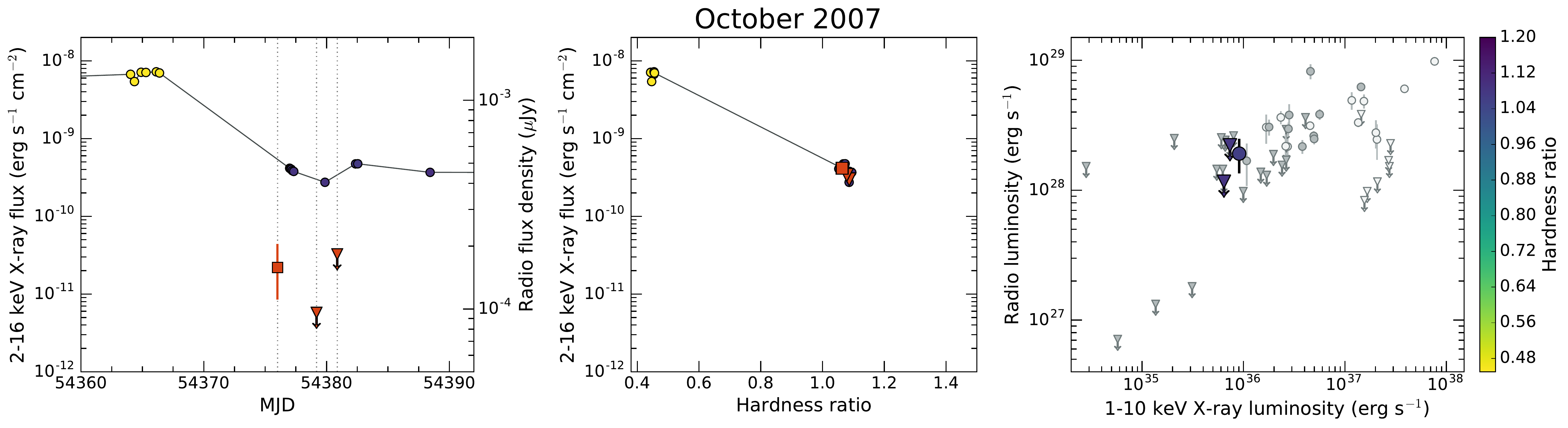}
\caption{Same as Fig. A.1, but for the October 2007 outburst.}
\vspace{-4mm}
\end{figure*}
\begin{figure*}
\centering 
\includegraphics[width=\textwidth]{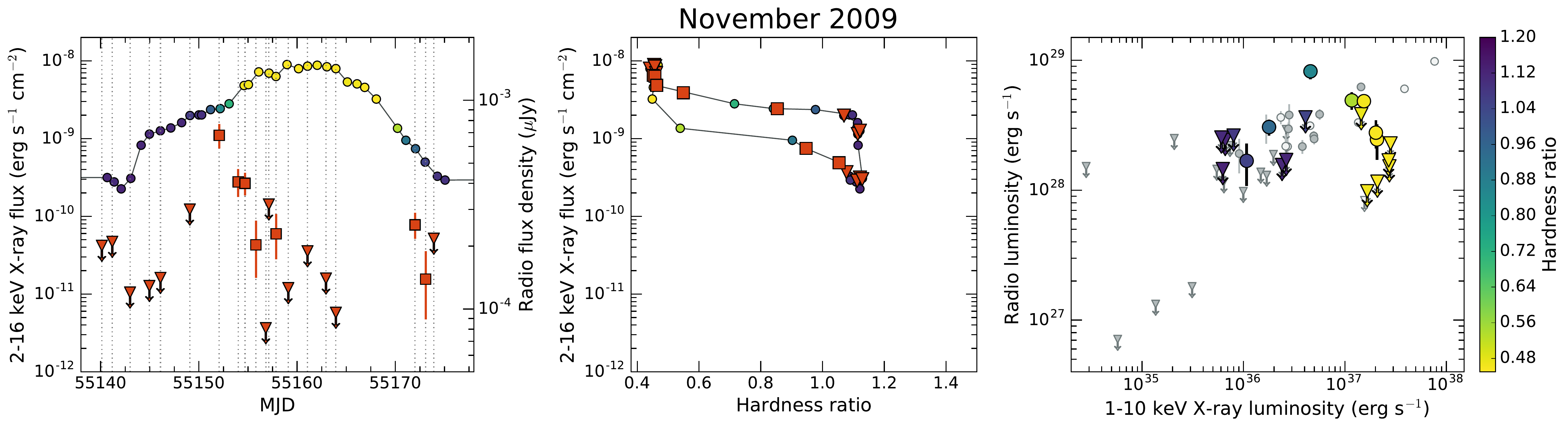}
\caption{Same as Fig. A.1, but for the November 2009 outburst and with radio observations taken with the VLA at 8.46 GHz, with the VLBA at 8.41 GHz and with the EVN at 5 GHz.}
\end{figure*}

\bsp	
\label{lastpage}
\end{document}